\documentclass[10pt,prd,preprintnumbers,floatfix,aps,nofootinbib,showpacs]{revtex4} 

\usepackage{epsfig}
\usepackage{url}
\usepackage{hyperref}

\usepackage[normalem]{ulem} 

\usepackage{latexsym}
\usepackage{epsfig}
\usepackage{amsmath}
\usepackage{amssymb}
\usepackage{graphicx}
\usepackage{verbatim}

\newcommand{\lp}{\left(}
\newcommand{\rp}{\right)}
\newcommand{\lb}{\left[}
\newcommand{\rb}{\right]}

\newcommand{\ba}{\begin{eqnarray}}
\newcommand{\ea}{\end{eqnarray}}
\newcommand{\be}{\begin{equation}}
\newcommand{\ee}{\end{equation}}

\newcommand{\half}{{1\over 2}}

\newcommand{\al}{\alpha}

\newcommand{\ga}{\gamma}

\newcommand{\Lag}{\mathcal{L}}

\newcommand{\mc}{\mathcal}
\newcommand{\ph}{\phantom{\al}}

\usepackage[usenames,dvipsnames]{xcolor}

\newcommand{\ud}[2]{^{#1}_{\phantom{#1} #2}}
\newcommand{\du}[2]{_{#1}^{\phantom{#1} #2}}

\newcommand{\enangle}[1]{\langle #1 \rangle} 

\usepackage[all]{xy} 

\begin{document} 

\title{{Shaken, not stirred:} kinetic mixing in scalar-tensor theories of gravity}

\author{Dario Bettoni}
\affiliation{Faculty of Physics, Israel Institute of Technology, \\ Technion City, 32000, Haifa, Israel}
\email{dario@physics.technion.ac.il}
\author{Miguel Zumalac\'arregui}
\affiliation{Institut f\"ur Theoretische Physik, Universit\"at Heidelberg \\ Philosophenweg 16, 69120 Heidelberg, Germany}
\email{zumalacarregui@thphys.uni-heidelberg.de}

\begin{abstract}
Kinetic mixing between the metric and scalar degrees of freedom is an essential ingredient in contemporary scalar-tensor theories. This often makes hard to understand their physical content, especially when derivative mixing is present, as it is the case for Horndeski action.
In this work we develop a method that allows to write a Ricci curvature-free scalar field equation and discuss some of the advantages of such rephrasing in the study of stability issues in the presence of matter, the existence of an Einstein frame and the generalization of the disformal screening mechanism. For quartic Horndeski theories, such procedure leaves, in general, a residual coupling to curvature, given by the Weyl tensor.
This gives rise to a binary classification of scalar-tensor theories into \emph{stirred theories}, for which the curvature can be substituted for, and \emph{shaken theories} for which a residual coupling to curvature remains. Quite remarkably, we have found that generalized DBI Galileons belong to the first class.
Finally, we discuss kinetic mixing in quintic theories for which non-linear mixing terms appears and in the recently proposed theories beyond Horndeski which display a novel form of kinetic mixing, in which the field equation is sourced by derivatives of the energy-momentum tensor. 
\end{abstract}

\date{\today}




\maketitle

\section{Introduction}

The evidence for accelerated stages of expansion in our universe's cosmological evolution has been increasing over the last years.
The first of such periods, cosmic inflation, would have occurred in the very early universe, giving birth to a flat, homogeneous and isotropic space-time, filled with thermalized radiation and imprinted with nearly scale-invariant and adiabatic perturbations. The second, late-time cosmic acceleration, reflects a much lower energy scale and only unfolds in the low redshift universe. 
Support for these phases of cosmological evolution comes from a number of complementary and increasingly precise probes that explore both the late and the early universe \cite{Weinberg:2012es,Ade:2013zuv,Ade:2013uln}. The implications of cosmic acceleration for fundamental physics will be further scrutinized with the next generation of experiments, like the Euclid satellite, the Dark Energy Spectroscopic Instrument and the Square Kilometer Array \cite{Amendola:2012ys,Levi:2013gra,Blake:2004pb}.

The simplest known mechanism for acceleration, a cosmological constant, can not satisfactory explain cosmic inflation without the introduction of a mechanism for its decay and a departure from scale invariance of the initial perturbations. Since inflation has to end, a more natural explanation is that it is produced by an additional dynamical degree of freedom, whose energy density eventually decays into dark matter and standard model particles \cite{Guth:1980zm,Linde:1981mu}. For what concerns the late-time acceleration, the cosmological constant is able to explain current observations for cosmic acceleration, but it suffers of many theoretical problems, which make the investigation of alternatives a compelling task \cite{Weinberg:1988cp,Sahni:1999gb,Martin2012}. A simple possibility to address the two phases of cosmic acceleration is the introduction of scalar degree(s) of freedom. Indeed, scalar fields are compatible with the symmetries of the cosmological space-time, can easily produce cosmic acceleration and occur naturally as limits of high energy theories of gravity. The search for models able to explain cosmic acceleration has triggered considerable interest in alternative gravitational theories (See \cite{Clifton:2011jh} for a recent review).

In particular, scalar-tensor (ST) theories of gravity have existed in the literature since the early sixties, when alternative theories were developed in parallel to increasingly precise tests of gravity in the Solar System \cite{Will:2014xja}. The interactions present in old-school Jordan--Brans--Dicke (JBD) theories \cite{Brans:1961sx} constitute the first generation of ST theories which was soon developed to a consistent framework for alternatives to Einstein gravity \cite{Fujii:2003pa,Faraoni:2004pi}. Recent developments in extra dimensions and massive gravity have also uncovered new theoretical frameworks which produce viable modifications of gravity: the generalization of the interactions found in the 5-dimensional Dvali--Gabadadze--Porrati braneworld model \cite{Dvali:2000hr} led to the proposal of Galileon ST field theories \cite{Nicolis:2008in}, which also arise naturally in the recently proposed dRGT ghost-free massive gravity \cite{deRham:2010kj} and bigravity \cite{Hassan:2011zd} in the limit in which gravity decouples (see \cite{Hinterbichler:2011tt,deRham:2014zqa} for reviews). While the aforementioned theories describe essentially different infrared physics, they are characterized by the same set of derivative interactions of the scalar field in the decoupling limit.

The generalization of these interactions to curved space-time \cite{Deffayet:2009mn} naturally leads to Horndeski's theory \cite{Horndeski:1974wa}, which was first proposed in the early seventies. This is the most general ST action in four dimensions whose variation produces second order equations of motion. It characterizes the second generation of ST theories and encompasses a large set of models proposed over the past years. As such, this theory has attracted considerable attention as a way to unify ST theories and study their phenomenology as applied to late time cosmology \cite{Amendola:2012ky,Gleyzes:2013ooa,Bellini:2014fua}, inflation \cite{Kobayashi:2011nu}, and local gravity tests \cite{Kase:2013uja}. Although this is the most general action whose variation produces equations of motion which are at most second order in derivatives, the completeness of Horndeski's theory as the master framework encompassing all viable ST theories has been recently challenged: examples of theories beyond Horndeski indicate the existence of a third generation of healthy ST theories \cite{Zumalacarregui:2013pma,Gleyzes:2014dya,Gleyzes:2014qga}.
Given this large number of different models and the need to check their viability, many attempts have been made in formulating model independent observables as to test general properties of ST theories \cite{Amendola:2012ky,Motta:2013cwa,Bellini:2014fua}, as well as in the search for connections between apparently different theories \cite{deRham:2010eu}.

This work provides a general procedure to investigate the properties of Horndeski's theory by using, as a starting point, the way the scalar and the metric degrees of freedom interact.
In fact, the aforementioned theories exhibit different degrees of kinetic interaction between the scalar and tensor degrees of freedom, a phenomenon known as \emph{kinetic mixing} or \emph{kinetic braiding} \cite{Deffayet:2010qz}. This property entangles the derivatives of the scalar and tensor field in the equations of motion, in a way that is unique to each theory. As we move from the simpler, JBD theories further into theories belonging to the second and third generation, the kinetic mixing becomes more intricate and new coupling structures appear. 
The idea is then to use the metric equations to remove all the instances of the curvature (which contains the second derivatives of the metric field) from the scalar field equation of motion. This procedure for \emph{covariant debraiding} greatly simplifies the study of the properties of the scalar degree of freedom.

We first illustrate how this procedure works for JBD ST theories and for cubic Galileons \cite{Deffayet:2010qz} and subsequently  we introduce the debraiding procedure for general quartic Horndeski theories. This case represents a completely new situation, since couplings between curvature and scalar derivatives appear at the level of the action, and we found several new aspects for this extension that are not present in simpler cases. In particular, this procedure can not be completed in general due to a coupling between the scalar field and the Weyl curvature tensor (the traceless part of the Riemann tensor), which is not algebraically determined by the metric equations of motion. The debraiding procedure can also introduce spurious solutions to the equations of motion. However, we show that it is always possible to select the physical branch of solutions.
We also found that in a specific subset of quartic theories, Dirac-Born-Infeld (DBI)-like theories, the debraiding procedure can be performed in an exact manner, and both the Weyl tensor and the spurious solutions are automatically eliminated. 
This result is expected, as such theories are equivalent to Einstein gravity via a disformal redefinition of the metric. 
These two different behaviours under the debraiding procedure suggest to classify ST theories according to the possibility to remove all curvature (\emph{stirred}) or not (\emph{shaken}).

The unbraided equations provide a new look into the properties of Horndeski theories: It unambiguously shows the couplings of the scalar field and sheds light to its behaviour within matter, as schematically shown in table \ref{table:mixing_classification}. Using our formalism we show that the DBI-like theories can present gradient instabilities in a radiation dominated universe with a sufficiently high pressure density, posing a serious challenge for the simplest among such theories. This instability can be easily avoided in non-DBI-like theories, for which the richer mixing structure can prevent the gradient instability. Finally, the unbraided equations allow one to generalize the disformal screening mechanism for scalar modifications of gravity \cite{Koivisto:2012za,Zumalacarregui:2012us}. This effect, which has been studied only for DBI-like theories in the Einstein frame allows the scalar field to evolve independently of the energy density if the scalar's time evolution is non-negligible and the energy density of matter is sufficiently large. In this work we show that this mechanism is present in a broader class of quartic Horndeski theories, in which it may work under a more relaxed set of assumptions.

We also show how the mixing structure acquires even more involved forms beyond quartic Horndeski theories. In quintic Horndeski theories the field equation has terms which are non-linear in the curvature, in the form of the Gauss-Bonnet scalar. Such non-linear mixing terms contain the square of the Weyl tensor and can not be debraided using our techniques. Theories beyond Horndeski introduce an even more subtle form of kinetic mixing, in which the scalar field is sourced by derivatives of the energy-momentum tensor (which can not be replaced using energy-momentum conservation in a covariant way). 

The paper is organized as follows. In section \ref{KMSCT} we will review the concept of kinetic mixing and covariant debraiding in old school and cubic ST theories. This study is extended to quartic theories in section \ref{section:quartic_mixing}, where we identify the mixing structures, apply the covariant debraiding program and comment on the new subtleties that appear. In section \ref{COKM} we will explore the consequences of such mixing, both in the general case and specializing to the specific case of quartic DBI Galileon \cite{deRham:2010eu}. In particular, we will discuss the relation between the coupling to the Weyl tensor and the existence of an Einstein frame, the stability of the theory and the generalization of the disformal screening mechanism.
In section \ref{section:beyond_quartic} we will extend some of the discussion to quintic theories and to theories beyond Horndeski, showing how new matter-scalar couplings appears in such models. Finally, in section \ref{CAO} we will discuss our results and draw the conclusions. 

We will work in four space-time dimensions, use a $-+++$ convention for the metric, set the speed of light and the reduced Planck constant to unity  $\hbar = c = 1$. Summation over repeated indices is assumed.

\begin{table}
 \begin{center}
\begin{tabular}{c  c  c  c }
	& Old-School & Horndeski & 3$^{\rm rd}$ generation \\[3pt] \hline\hline  \\[-2.ex]
Examples & JBD/$f(R)$ & covariant Galileons & covariantized Galileons \\[3pt] \hline \\[-2.ex] 
Ostrogradski Degeneracy & No $\partial^2\phi$ in $\Lag$ & 2$^{\rm nd}$-order eqs. & {Implicit constraints} \\[3pt] \hline \\[-2.ex]
	& algebraic  & algebraic & derivative \\[3pt]	  
Kinetic Mixing	  & $T$ & $T_{\mu\nu}\phi^{;\mu\nu}$,  $T_{\mu\nu}\phi^{,\mu}\phi^{,\nu}$  & $\nabla(T_{\mu\nu}\phi^{,\mu}\phi^{,\nu})$, $\nabla T$,...  \\
	  &  & $W_{\mu\alpha\nu\beta}\phi^{;\mu\nu}\phi^{,\alpha}\phi^{,\beta}$,... & 
\\[3pt] \hline \\[-2.ex]
Formal Invariance \\ under metric redefinitions & $C(\phi)g_{\mu\nu}$ & $C(\phi)g_{\mu\nu}+D(\phi)\phi_{,\mu}\phi_{,\nu}$ & $D(X,\phi)\phi_{,\mu}\phi_{,\nu}+\cdots$ \\  \hline\hline
\end{tabular}
\end{center}
\caption{Three generations of ST theories and some of their theoretical properties. The first line describes how the Ostrogradski degeneracy \cite{Woodard:2006nt} is avoided as we move towards more complex ST theories. In the second line, the typical scalar-matter couplings that appear after the debraiding procedure is carried out. Finally the third line, shows which metric redefinition leaves the action formally invariant, amounting to a redefinition of the functions that specify the model.
\label{table:mixing_classification}}
\end{table}
 
\section{Kinetic mixing: definition and simple cases}
\label{KMSCT}

In the introduction we have pointed out how contemporary ST theories can show a very complex mixing of their degrees of freedom in a way that is specific to the theory at hand. In particular, we have stressed how the couplings between metric derivatives and scalar field in the action will lead to second derivatives of one field to act as source for the other and to couplings between them, resulting in a very complex coupled dynamical system. In general, this mixing will not only complicate the numerical solution of the equations, but will also obscure their physical interpretation.

One interesting way to simplify the scalar field equations is what we will call the \emph{covariant debrading} procedure. This basically amounts to identify the scalar-curvature couplings that appear in the scalar equation, and use contractions of the metric equations with the scalar field derivatives to trade those for terms which depend on the scalar and matter fields. 
The outcome of this procedure will be a scalar equation of motion that depends on the scalar and matter fields and whose only second derivatives are those of the scalar field.

There are several reasons for pursuing this idea. First of all, an unmixed equation of motion for the scalar field makes clear the interaction structure of the theory and the couplings between matter and the scalar field. This allows to use the debraided equation  to study the stability of a given theory without the need to take into account also the metric equations. The kinetic mixing properties of a theory also determine its phenomenology; for example, kinetic mixing is necessary for any model to have a variable effective gravitational constant in cosmological scenarios \cite{Bellini:2014fua}. Another reason is that this procedure allows a classification of different models depending on their matter and self-interactions. In fact, we will see that these are unique features of any given model and can help to distinguish between theories that are not equivalent via field redefinitions. Finally, we notice that this procedure has the additional advantage of being fully non-linear and covariant (not relying on a specific expansion or choice of background) and of using the Jordan frame matter stress-energy tensor which is covariantly conserved. This is simpler than rewriting the theory in the Einstein frame, which is not possible in general and leads to the energy-momentum being sourced by the scalar explicitly.
 
In this paper we will mainly work with Horndeski's theory \cite{Horndeski:1974wa} in its modern formulation \cite{Deffayet:2011gz}, described by the following action
\begin{equation}
S_H[g_{\mu\nu},\phi] = \int d^4 x \sqrt{-g} \sum_{i=2}^4 \Lag_i\,,
\end{equation}
 where
\begin{eqnarray}
\Lag_2 &=&   G_2(X,\phi)\,, \label{LH2} 
 \\[5pt]
\Lag_3 &=& G_3(X,\phi) \Box\phi 
\,, \label{LH3}
 \\[5pt]
\Lag_4 &=& G_4(X,\phi) R + G_{4,X}\lp (\Box\phi)^2 - (\phi_{;\mu\nu})^2 \rp\,, \label{LH4}
\\[5pt]
\Lag_5 &=& G_5(X,\phi) G_{\mu\nu}\phi^{;\mu\nu}  - \frac{1}{6}G_{5,X}\lp (\Box\phi)^3
- 3\Box\phi(\phi_{;\mu\nu})^2 + 2(\phi_{;\mu\nu})^3 \rp \,, \label{LH5}
\end{eqnarray}
are the \emph{quadratic}, \emph{cubic}, \emph{quartic} and \emph{quintic} Lagrangian respectively.%
\footnote{These names have historic origin. They refer to the power of the field in flat-space Galileons, for which $G_1 \propto \phi$, $G_2, G_3\propto X$, $G_4,G_5 \propto X^2$.}
Here $X\equiv -\frac{1}{2}g^{\mu\nu}\phi_{,\mu}\phi_{,\nu}$, and $\Box\phi = \phi\ud{;\mu}{;\mu}$, $(\phi_{;\mu\nu})^n = \phi\ud{;\alpha_n}{\alpha_1}\cdots\phi\ud{;\alpha_{n-1}}{;\alpha_n}$ denote contractions of the field's second derivatives.
As we have discussed in the introduction this theory has attracted considerable attention in recent years as a way to unify ST theories and study their phenomenology and hence represents the best framework for investigating the debrading procedure. 

We will introduce the essential features of kinetic mixing by presenting results for  JBD theories and cubic Galileons (the simplest of Horndeski's theories) in which all the basic features are already present. Quartic Horndeski theories will be presented separately in section \ref{section:quartic_mixing}  and \ref{COKM} while the novel kinetic mixing features introduced in quintic and non-Horndeski theories will be briefly discussed in section \ref{section:beyond_quartic}.

\subsection{Old-school scalar-tensor theories}\label{section:old_school_mixing}

Let us start exploring the issue of kinetic mixing in JBD theories of gravity. Here we focus on a theory described by a coupling between the field and the Ricci scalar, a canonical kinetic term and a potential for the field:%
\footnote{This is not the most general formulation of an old-school ST theory but such theory can always be mapped to this form with suitable field redefinitions \cite{Flanagan:2004bz,Sotiriou:2007zu}}
\begin{equation}
 S_{\rm JBD} = \int d^4 x\sqrt{-g}\left( \frac{M_p^2}{2}C(\phi)^2R + X - V(\phi) + \mathcal L_m\right)\,,
\end{equation}
The dynamics of the above theory is described by the metric equation
\begin{equation}\label{eq:jbd_metric}
 C^2 G_{\mu\nu} + (g_{\mu\nu}\Box C^2 - C^2_{;\mu\nu}) = 
 \frac{1}{M_p^2} \left( T^{(m)}_{\mu\nu} + \phi_{,\mu}\phi_{,\nu} +  g_{\mu\nu}(X-V) \right)\,,
\end{equation}
where the energy-momentum tensor is defined as $T_{\mu\nu}=\frac{-2}{\sqrt{-g}}\frac{\delta(\sqrt{-g}\mathcal L_m)}{\delta g^{\mu\nu}}$, and the scalar field equation
\begin{equation}\label{eq:jbd_field}
 \Box\phi - V^\prime + CC^\prime M_p^2 R = 0\,.
\end{equation}
The kinetic mixing is reflected in the fact that the kinetic terms of the metric ($\sim R_{\mu\nu}$) and of the field ($\phi_{;\mu\nu} \sim \nabla\nabla C$) appear in both equations. It is possible to debraid the scalar field equation by taking the trace of the metric equations and substituting it in (\ref{eq:jbd_field})

The debraided field equation has the following structure
\begin{equation}
 \underbrace{\left(1+6M_{Pl}^2C^{\prime 2}\right)\Box\phi}_\text{renormalized kinetic term} 
\!\!\! - V^\prime 
+ \underbrace{4\frac{C_{,\phi}}{C}V -2X \frac{C_{,\phi}}{C}\left(6 M_{Pl}^2(C^{\prime 2}+CC_{,\phi\phi})+1\right)}_\text{additional terms}
  = \underbrace{\frac{C_{,\phi} }{C} T}_\text{matter coupling}\,,
\end{equation}
and contains no second derivatives of the metric. This simple example already reveals some of the debraiding features that will occur on more general theories:
\begin{enumerate}
\item There is an explicit coupling to matter, leading to an environment-dependent effective potential. In this simple case it is proportional to the trace of the energy-momentum tensor $T$, as could be anticipated from coupling between the Ricci scalar and the scalar field function $C$ in eq. (\ref{eq:jbd_field}). Notice that this coupling is \emph{one way}, in the sense that a minimally coupled matter source will still have the matter stress-energy tensor conserved.
\item The coefficient of the second derivative term gets renormalized by a function of the field, showing how kinetic mixing can affect the stability properties of the scalar field equation.
In this case the coefficient is strictly positive and hence no instabilities can be dynamically generated. However, this will not generally be true for more complex theories.
\item New terms not involving second derivatives appear in the equation,  coming from the contraction of the first derivative terms in the metric equations. In particular, the potential term for the field is modified and a new term involving first derivatives of the field appears.
\end{enumerate}
It is worth stressing that matter is minimally coupled and therefore the energy-momentum tensor is covariantly conserved $\nabla_\mu T^{\mu\nu}_{(m)}=0$. This would not be true if similar results were obtained by expressing the theory in the Einstein frame by a redefinition of the metric. 

\subsection{Cubic Horndeski theories} \label{section:cubic_mixing}

Cubic Horndeski theories (\ref{LH3}) are characterized $G_4=M_{Pl}^2/2$, $G_5=0$ and generic functions $G_3$ and $G_2$. As we will see, they contain richer forms of kinetic mixing in curved space-time. Here we will explore their behaviour for the simplest non-trivial example, the cubic Galileon:
\begin{equation}
 S_{\rm CG} = \int d^4x\sqrt{-g}\left( \frac{M_p^2}{2}R + X + \frac{X}{\Lambda^3}\Box\phi\right)\,.
\end{equation}
The second derivatives of the field present in the last term produce a coupling with the affine connection, which in turn introduces a term involving the curvature in the field equation
\begin{equation}\label{eq:cubic_curvature_coupling}
 \frac{\delta S}{\delta\phi} = \Lambda^{-3}\phi_{,\mu}R^{\mu\nu}\phi_{,\nu}+\;\text{terms without curvature}\,,
\end{equation}
(one can alternatively see the the emergence of the Ricci tensor through the anti-commutation of covariant derivatives, which appear anti-symmetrically in the equations of motion). Cubic Horndeski theories are known as \emph{kinetic gravity braiding} (KGB) \cite{Deffayet:2010qz,Pujolas:2011he} for this reason.

Just as in the JBD case, it is possible to use contractions of the metric equations to solve for the curvature coupling in (\ref{eq:cubic_curvature_coupling}). The only difference is that one has to contract with both the metric and $\phi_{,\mu}\phi_{,\nu}$.
The resulting debraided field equation reads
\begin{eqnarray}
&& \underbrace{\left(\big(1-\frac{2X^2}{M_p^2\Lambda^6}\big)g^{\mu\nu}-\frac{4X}{M_p^2\Lambda^6}\phi^{,\mu}\phi^{,\nu} \right)\phi_{;\mu\nu}}_\text{renormalized kinetic term}
-\underbrace{\frac{4X^2}{M_p^2\Lambda^3}}_\text{extra terms}
\nonumber \\ 
&&+\underbrace{\frac{1}{\Lambda^3\phantom{_p}} \Big[ (\Box\phi)^2 -\phi_{;\mu\nu}\phi^{;\mu\nu}\Big]}_\text{higher derivative interactions}
-\underbrace{\frac{1}{M_p^2\Lambda^3}\left(\phi_{,\mu}T^{\mu\nu}\phi_{,\nu} +T X\right)}_\text{coupling to matter}
=0\,.
\end{eqnarray}
We note the following features:
\begin{enumerate}
 \item The coupling to matter has two contributions, a conformal one (proportional to the trace $T$ and weighted by $X$) and a disformal one given by the contraction of the energy-momentum tensor with $\phi_{,\mu}\phi_{,\nu}$. The disformal part is particularly interesting, as it indicates that radiation would have non-trivial effects on the field in this type of theories.
 \item The kinetic term is renormalized due to the braiding by both conformal $\propto g^{\mu\nu}$ as well as disformal $\propto \phi^{,\mu}\phi^{,\nu}$ terms. Unlike in the old-school case, the corrections are not positive definite any more.
 \item There appears a \emph{Galileon term} constructed out of anti-symmetric, non-linear second derivative contractions. This term is not renormalized in the unbraided form of the equations. This type of terms are responsible for the Vainshtein screening mechanism in cubic theories that allows these theories to fit local gravity tests.
\end{enumerate}
The authors of \cite{Deffayet:2010qz} used the debraided equations to study causality and stability of KGB theories. In what follows we will extend the same program to quartic Horndeski theories and discuss the new features and subtleties that appear.

\section{Kinetic mixing in Quartic Horndeski theories} 
\label{section:quartic_mixing}

In this section we extend the debraiding formalism to Quartic Horndeski theories defined by the fixing $G_5(\phi,X)=0$ and $G_3(\phi,X)=0$ while leaving the other two functions arbitrary.%
\footnote{While the second choice is just for practical use, as it simplifies the calculations without significantly altering the analysis, the first one is more important. In fact, taking $G_5(\phi,X)\neq0$ would introduce significant deviations as we will discuss later on.}

These theories, defined in (\ref{LH4}), introduce a range of new kinetic mixing terms which arise in the field equation via both the cancellation of higher derivatives (due to the anti-symmetric structure, as in cubic theories) and the direct coupling between the field and the Ricci scalar in the action: $G_4(X,\phi) R$.%
\footnote{Note that the dependence of $X$ generates derivatives of $R$ in the field equation, which cancel exactly with counter terms stemming from the pure field part \cite{Deffayet:2009wt}. This follows from the defining property of Horndeski theories, namely the absence of higher than second derivatives in the Euler--Lagrange variation.}
The metric and scalar field equations are reported in appendix \ref{section:full_eqs_quartic}, while here we focus on the mixing terms which appear in the equation of motion with the following structure:
\begin{eqnarray}
 \frac{\delta\mathcal L_4}{\delta\phi} &=& 
 - 2G_{4,X}G^{\alpha\beta}\phi_{;\alpha\beta}
 +2G_{4,XX}\phi^{,\alpha}\phi^{,\beta}\left(2\phi\du{;\alpha}{;\lambda}R_{\lambda\beta} 
 + \phi^{;\mu\nu}R_{\mu\alpha\nu\beta} - \Box\phi R_{\alpha\beta} - \frac{R}{2}\phi_{;\alpha\beta}\right)
 \nonumber \\
 && + G_{4,\phi}R - 2G_{4,X\phi}R_{\alpha\beta}\phi^{,\alpha}\phi^{,\beta} +\; \text{terms without curvature}\,.
\end{eqnarray}
We note two distinguishing features that did not appear in lower order theories. First, there are second derivatives of the scalar field multiplied by Ricci curvature. Therefore, using the metric equations will introduce new products of field derivatives into the equations, which will in general fail to be linear in the second time derivatives and hence may introduce spurious solutions to the equations of motion.
Second, the derivatives of the field also couple to the full Riemann tensor. This term can be rewritten in terms of the Weyl tensor as
\begin{equation}
 R_{\mu\nu\alpha\beta} = W_{\mu\nu\alpha\beta}+g_{\mu[\alpha}R_{\beta]\nu}-g_{\nu[\beta}R_{\alpha]\mu}-\frac{1}{3}Rg_{\mu[\alpha}g_{\beta]\nu}\,, \label{eq:weyl_def}
\end{equation}
where the square brackets stands for antisymmetrization of the $n$ encompassed indices with weight $1/n!$ . This is a necessary step in order to split the Riemann tensor into its trace part, solvable from the metric equations and into its traceless part, which cannot be solved for using contractions of the metric equations.
Besides introducing new interesting features, both aspects represent an obstruction to the debraiding process. The spurious solutions are a technical complication that can be surpassed, as we will explain in section \ref{sec:spurious_sol}. On the other hand, the Weyl coupling is not devoid of physical meaning and its consequences are explored in section \ref{sec:weyl}.

In the case under investigation, where $G_4=G_4(\phi,X)$ and $G_2=G_2(\phi,X)$, the field Euler--Lagrange equation can be written in the following compact form
\begin{equation}
L^{\mu\nu}\phi_{;\mu\nu}+V+P^{\mu\nu\alpha\beta}\phi_{;\mu\nu}\phi_{;\alpha\beta} +Q^{\mu\nu\alpha\beta\rho\sigma}\phi_{;\mu\nu}\phi_{;\alpha\beta}\phi_{;\rho\sigma}=0\,, \label{eq:quartic_debraided}
\end{equation}
with
\begin{eqnarray}
\nonumber
L^{\mu\nu}&=&\left[G_{2,X}+G_{4,XX}\left(G_{\alpha\beta} \phi^{,\alpha}\phi^{,\beta} -\frac{4}{3}XR\right)\right]g^{\mu\nu}-\left(G_{2,XX}-\frac{1}{3}G_{4,XX}R\right)\phi^{,\mu}\phi^{,\nu}\\
&-&2\left(G_{4,X}+G_{4,XX}\right)G^{\mu\nu}+2G_{4,XX}\left(\phi^{,\alpha} G_{\alpha}{}^\mu\phi^{,\nu}
+\phi^{,\alpha}\phi^{,\beta} W^\mu{}_\alpha{}^\nu{}_\beta\right)\,, \label{eq:debraided_L_term}\\
\nonumber
P^{\mu\nu\alpha\beta}&=&\left(3G_{4,\phi X}-2G_{4,\phi XX}\right)\left(g^{\mu\nu}g^{\alpha\beta}-g^{\mu\alpha}g^{\nu\beta}\right)+2G_{4,\phi XX}\left(2\phi^{,\mu}\phi^{,\alpha} g^{\nu\beta}-2\phi^{,\mu}\phi^{,\nu} g^{\alpha\beta}\right)\,,\\
&&\\
Q^{\mu\nu\alpha\beta\rho\sigma}&=&G_{4,XX}\left(g^{\mu\nu}g^{\alpha\beta}g^{\rho\sigma}-3g^{\mu\nu}g^{\alpha\rho}g^{\beta\sigma}+2g^{\mu\sigma}g^{\nu\alpha}g^{\beta\rho}\right) \nonumber \\
&-&G_{4,XXX}\left(2\phi^{,\mu} g^{\nu\alpha}g^{\beta\rho}\phi^{,\sigma}-2\phi^{,\mu} g^{\nu\alpha}g^{\rho\sigma}\phi^{,\beta}+\phi^{,\mu}\phi^{,\nu}\left(g^{\rho\sigma}g^{\alpha\beta}-g^{\rho\alpha}g^{\sigma\beta}\right)\right)\,,\\
V&=&G_{2,\phi}-2XG_{2,\phi X}-4G_{4,\phi X}\phi^{,\alpha}G_{\alpha\beta}\phi^{,\beta}+G_{4,\phi}R\,.
\end{eqnarray}
where the first term is linear in second derivatives, the second one is a potential term that depends at most on first derivatives of the field while the last two terms, despite being respectively quadratic and cubic in derivatives, are linear in second time derivatives and non-linear only in mixed spatial derivatives.

As it has been done in the previous cases we can eliminate the curvature-field couplings with suitable contractions of the metric equations with scalar field derivatives. In this case the structure is more involved and hence we describe it in a schematic way, leaving the full expressions to appendix \ref{section:full_eqs_quartic}. The debraided field equation is
\begin{eqnarray}
 \tilde L^{\mu\nu}\phi_{;\mu\nu}+\tilde{V}+\mathcal{Q}_T T+\mathcal{Q}_{\enangle{T}}\phi^{,\beta}T_{\alpha\beta}\phi^{,\alpha}+\left(\tilde{P}^{\mu\nu\alpha\beta}+K^{\mu\nu\alpha\beta}\right)\phi_{;\mu\nu}\phi_{;\alpha\beta} \nonumber
 &&\\
 +\left(H^{\mu\nu\alpha\beta\rho\sigma}+\tilde Q^{\mu\nu\alpha\beta\rho\sigma}\right)\phi_{;\mu\nu}\phi_{;\alpha\beta}\phi_{;\rho\sigma}&=&0
 \label{eq:unbraided_schematic}\,,
\end{eqnarray}
with
\begin{eqnarray}
\nonumber
\tilde L^{\mu\nu}&=&\left(\mathcal G_0 
+\mathcal G_{\enangle{T}}\phi^{,\beta}T_{\alpha\beta}\phi^{,\alpha}
+\mathcal G_T T\right)g^{\mu\nu} 
+\left(\mathcal S_0 +\mathcal S_{\enangle{T}}\phi^{,\beta}T_{\alpha\beta}\phi^{,\alpha}
+\mathcal S_T T\right)\phi^{,\mu}\phi^{,\nu}\\
&+&\mathcal C_{\enangle{T}}\phi^{,\sigma} T_\sigma{}^\mu\phi^{,\nu}+\mathcal C_T T^{\mu\nu} 
+\mathcal C_W\phi^{,\alpha}\phi^{,\beta} W^\mu{}_\alpha{}^\nu{}_\beta\,, \label{eq:4_unbraided_open_indices}
\\ 
\tilde P^{\mu\nu\alpha\beta}&=&\mathcal V_{4B}\left(g^{\mu\nu}g^{\alpha\beta}-g^{\mu\alpha}g^{\nu\beta}\right)+\mathcal V_{4D}\left(\phi^{,\mu}\phi^{,\alpha} g^{\nu\beta}-\phi^{,\mu}\phi^{,\nu} g^{\alpha\beta}\right)\,,\\
K^{\mu\nu\alpha\beta}&=&\mathcal{W}_{D2}\left(\phi^{,\mu}\phi^{,\nu}\phi^{,\alpha}\phi^{,\beta}+4X^2g^{\mu\nu}g^{\alpha\beta}+4X\phi^{,\mu}\phi^{,\nu} g^{\alpha\beta}\right)\,,\\
\nonumber
\tilde Q^{\mu\nu\alpha\beta\rho\sigma}&=&
\mathcal V_{5g} \left(g^{\mu\nu}g^{\alpha\beta}g^{\rho\sigma}-3g^{\mu\nu}g^{\alpha\rho}g^{\beta\sigma}+2g^{\mu\sigma}g^{\nu\alpha}g^{\beta\rho}\right)\\
&-&\mathcal V_{5X} \left(2\phi^\mu g^{\nu\alpha}g^{\beta\rho}\phi^{,\sigma}-2\phi^{,\mu} g^{\nu\alpha}g^{\rho\sigma}\phi^{,\beta}+\phi^{,\mu}\phi^{,\nu}\left(g^{\rho\sigma}g^{\alpha\beta}-g^{\rho\alpha}g^{\sigma\beta}\right)\right)\,,\\
H^{\mu\nu\alpha\beta\rho\sigma}&=&
\mathcal W_1\left(\left(g^{\mu\nu}g^{\alpha\beta}-g^{\mu\alpha}g^{\nu\beta}\right)\left(\phi^{,\rho}\phi^{,\sigma}+2Xg^{\rho\sigma}\right)\right. \nonumber \\
\nonumber
&+&\left.3\left(\phi^{,\mu}\phi^{,\beta} g^{\nu\alpha}-g^{\mu\nu}\phi^{,\alpha}\phi^{,\beta}-X\left(g^{\mu\nu}g^{\alpha\beta}-g^{\mu\alpha}g^{\nu\beta}\right)\right)\right)g^{\rho\sigma}\\
&+& \mathcal W_2 
\left(\phi^{,\mu}\phi^{,\beta} g^{\nu\alpha}-g^{\mu\nu}\phi^{,\alpha}\phi^{,\beta}-X\left(g^{\mu\nu}g^{\alpha\beta}-g^{\mu\alpha}g^{\nu\beta}\right)\right)\left(\phi^{,\rho}\phi^{,\sigma}+2Xg^{\rho\sigma}\right)\,. \qquad
 \label{eq:4_unbraided_higher_power}
\end{eqnarray}
Here the coefficients $\tilde V, \mathcal G_i, \mathcal S_i, \mathcal C_i, \mathcal V_i, \mathcal W_i, \mathcal Q_i$ depend on $\phi$ and $X$ through $G_4$, $G_2$ and their partial derivatives and are fixed once a specific model is chosen. In table \ref{table:mixing_sumary} we schematically report the coefficient structure for three models of quartic Horndeski, while the general expressions can be seen in appendix \ref{sec:debraided_quartic_coefficients}.

After the debraiding process some of the terms in the debraided equation take a form analogous to the unbraided one but with ``renormalized'' structure coefficients. Extra terms which couple matter to the scalar field are also introduced. 
In this regard, it is interesting to note that these coefficients (see appendix \ref{section:full_eqs_quartic}) have a common denominator structure
\begin{equation}
\mathcal{X}_i\sim \left(G_4-2X G_{4,X}\right)^{-n}\left(G_4-X G_{4,X}\right)^{-m}\,.
\end{equation}
In particular, the coefficients can become singular for certain values of the field. However, the first factor is inversely proportional to the effective Planck mass, defined as the coefficient of the second time derivative in the graviton propagation equation. In homogeneous and isotropic backgrounds, and restricting to quartic theories, this has been shown to be \cite{Bellini:2014fua}
\begin{equation}
 M_{eff}^2 = 2(G_4 - 2X G_{4,X} )\,. 
 \label{Eff_Planck_Mass}
\end{equation}
This equation can be covariantized right away, suggesting its validity on general backgrounds.
This dependence is consistent with the fact that the debraiding procedure substitutes the curvature terms with the energy-momentum tensor via the metric equations and hence it suppresses the new terms by a Planck mass factor.
Therefore the singularity of the unbraided equations is related to a physical singularity, as the coefficient of the graviton kinetic term vanishes.
Moreover, requiring the positivity of the effective Plank mass also implies that the other factor in the denominator will never become singular as long as $G_4$ is positive. 

The first term in equation (\ref{eq:unbraided_schematic}), as can be seen from (\ref{eq:4_unbraided_open_indices}), is linear in second derivatives and is of particular interest as it contains the couplings between matter fields and second derivatives of the scalar field. This represents a new type of contributions with respect to previous cases and is important for the stability of these models. As we will discuss below, depending on the nature and evolution of matter fields, instabilities may occur for certain theories in the presence of matter. We point out again the presence of the Weyl term, indicating that there is still a residual dependence on the second derivatives of the metric whose physical meaning will be discussed in section \ref{sec:weyl}.

The $Q_i$ terms contains other couplings between the scalar field and matter which involve at most first derivatives of the scalar field, while the other terms are non-linear derivative interactions, responsible, for example, for the Vainishtein screening mechanism. However, among those, the terms proportional to $H^{\mu\nu\alpha\beta\rho\sigma}$, $K^{\mu\nu\alpha\beta}$ can be quadratic in second time derivatives on non-trivial backgrounds. This dependence may introduce spurious solutions, which can nonetheless be distinguished from the physical ones (see section \ref{sec:spurious_sol}).

\subsection{Spurious solutions and their avoidance.}\label{sec:spurious_sol}

The covariant debraiding procedure introduces unphysical solutions in the field equation. Contractions of the metric equations with $\phi_{;\mu\nu}$ and $\phi\du{;\mu}{;\lambda}\phi_{,\nu}$ (necessary to the covariant debraiding process) lead to the introduction of quadratic powers of second time derivatives in eq. (\ref{eq:unbraided_schematic}). Therefore, two branches exist for the solutions, the physical one and a spurious mode, which has to be disregarded.%
\footnote{A cartoon example of this: If our original equation has the form $f(x)=a$, taking the square will introduce a spurious solution for $x$, corresponding to $f(x)=-a$. We thank J. Beltr\'an for helping us clarify the introduction of spurious solutions.}

The debraiding procedure is equivalent to summing the field equation and the combination
\begin{equation}
 \mathcal M \equiv \left( c_0 g^{\mu\nu} + c_1 \phi^{,\mu}\phi^{,\nu} + c_2 \phi^{;\mu\nu} + c_3 \phi^{,\mu}\phi^{;\nu\alpha}\phi_{,\alpha}\right) \left(\mc E_{\mu\nu} - T_{\mu\nu}\right)=0\,,
\end{equation}
where $\mc E_{\mu\nu} = \frac{1}{\sqrt{-g}}\frac{\delta \left(\sqrt{-g}\mc L_4\right)}{\delta g^{\mu\nu}}$ and $c_0$-$c_3$ are chosen to cancel the braiding terms in the field equation. 
The problematic $\ddot\phi^2$ terms appear in the structure term $H^{\mu\nu\alpha\beta\rho\sigma}$ and $K^{\mu\nu\alpha\beta}$ in eq. (\ref{eq:unbraided_schematic}) as a consequence of the contraction of second derivatives of the scalar field with the metric equations that also contain such terms that lacks of the required antisymmetric structure required to avoid non-linear terms.
Note that the coefficient of $\ddot\phi^2$ involves contractions of spatial derivatives and hence solutions on simple geometries (e.g. FRW) will not display the complications associated with the spurious solutions. 

We can use a trick to help us pick up the correct branch for the solution.
Instead of adding $\mathcal M$ to the field equation, we can deal instead with
\begin{equation}
 \frac{\delta \mc L}{\delta \phi} + \varepsilon \mathcal M = 0\,.
\end{equation}
The above equation allows one to regard the debraiding procedure as a continuous deformation of the field equation
depending on a parameter $\varepsilon$: it is now possible to interpolate between the original field equation $\varepsilon = 0$ (which is linear in second time derivatives) and the unbraided field equation $\varepsilon = 1$. 
Once a time direction has been chosen (e.g. through an ADM decomposition or an explicit choice of coordinates), one can schematically study the field equation
\begin{equation}\label{eq:ddphi_squared_schematic}
 \varepsilon \ddot\phi^2 + B \ddot\phi + C = 0\,,
\end{equation}
where $\varepsilon$ has been left explicit only in the term quadratic in second time derivatives. This form is guaranteed by the fact that all non-linearities in second time derivatives are introduced by the unbraiding terms, and are therefore linear in $\varepsilon$. The existence of real solutions will be determined by the condition $B^2 - 4\varepsilon C>0$.%
\footnote{It is theoretically possible that the debraiding equations have no real solution, while the original ones do (if $B^2 < 4C$). In that case one would need to diagonalize the system of differential equations in some other way in order to numerically integrate the dynamics.}
In that case there will be two solutions
\begin{equation}
 \ddot\phi = \frac{1}{2\varepsilon}\left(-B\pm \sqrt{B^2-4\varepsilon C}\right) 
  = \frac{B}{2\varepsilon}(1\pm 1) \mp \frac{C}{B} + \mc O (\epsilon)\,,
\end{equation}
where the last equality entails an expansion on $\varepsilon$. We can identify the unphysical branch as the one associated with the $+$ sign, as it is not mapped to a solution of the original equation when $\varepsilon\to 0$. As the physical solution should depend continuously on $\varepsilon$, the solution associated to the minus sign is the physical one while the plus sign leads to a spurious solution which is not originaly present.

Summarizing, one can choose the physical solution of the unbraided equations after choosing a time coordinate by writing the field equation in the form (\ref{eq:ddphi_squared_schematic}), setting $\varepsilon = 1$ and integrating in time $\ddot\phi$ using the solution with the negative sign.

\begin{table}[t]
\begin{center}
 \begin{tabular}{l c c c c c c c c c c c c c c c c}
$\frac{2}{M_{Pl}^2}G_4(\phi,X)$& $\mathcal{G}_T$ & $\mathcal{G}_{\enangle{T}}$ & $\mathcal{S}_T$ & $\mathcal{S}_{\enangle{T}}$ & $\mathcal{C}_T$ & $\mathcal{C}_{\enangle{T}}$ & $\mathcal{Q}_T$ & $\mathcal{Q}_{\enangle{T}}$ & $\mathcal{C}_W$ & $\mathcal{V}_{4D}$ & $\mathcal{V}_{4B}$ & $\mathcal{V}_{5D}$ & $\mathcal{V}_{5B}$ & $\mathcal{W}_{1}$ & $\mathcal{W}_2$ & $\mathcal{W}_{D2}$
\\ \hline\hline\\
$\sqrt{1-2X/\Lambda^4}$ (\ref{eq:dbi_unbraided_Kin})               & $0$ & $0$ &  $0$ & $0$& $\checkmark$ & $0$ & $0$ & $0$  & $0$ & $0$ &  $0$ & $0$ & $0$ & $0$ & $0$ & $0$ \\
$\sqrt{1-2A(\phi)X/\Lambda^4}$ (\ref{eq:dbi_unbraided})        & $0$ & $0$ &  $0$ & $0$& $\checkmark$ & $0$ & $\checkmark$ & $\checkmark$   & $0$ & $\checkmark$ &  $\checkmark$ & $0$ & $0$ & $0$ & $0$ & $0$\\
$\frac{(1+(X/\Lambda^4)^n)}{(1+(X/\Lambda^4)}$ \cite{Amendola:2014wma}               &  $\checkmark$ &  $\checkmark$ &   $\checkmark$ &  $\checkmark$ &  $\checkmark$ & $\checkmark$ & $0$ & $0$            &  $\checkmark$ & $0$ &  $0$ &  $\checkmark$ &  $\checkmark$&  $\checkmark$&  $\checkmark$& $0$
\\ \hline\hline
\end{tabular}
\end{center}
\caption[Debraided coefficients for three models of quartic Horndeski action.]{Debraided coefficients scheme for three models of quartic Horndeski action. The ticks indicate which one of the debraided coefficient is present for each theory.
\label{table:mixing_sumary}}
\end{table}

\subsection{DBI-like Galileons}
\label{sec:DBI_galileon}

In this section we are going to apply the debraiding method described in the previous section to a specific set of models for which the debraided equations are greatly simplified. We will then use these models as a reference when discussing the applications of the debrading method. Quite remarkably, these models are a generalization of the quartic Dirac-Born-Infeld (DBI) Galileon \cite{deRham:2010eu}, in which
\begin{equation}
 G_4(\phi,X) = \frac{M_p^2}{2}\sqrt{1 - 2A(\phi)\frac{X}{\Lambda^4}}\,,
\end{equation}
where $\Lambda$ is a new mass scale, while $G_2$ is left generic.%
\footnote{We could have added a field dependent Planck mass, but this would not modify most of the results.}

The debraided field equation for this theory has the remarkably simple form:
\begin{equation}
 \tilde L^{\mu\nu}\nabla_\mu\nabla_\nu\phi+\tilde{V}+\mathcal{Q}_T T+\mathcal{Q}_{\enangle{T}}\phi^{,\beta}T_{\alpha\beta}\phi^{,\alpha}+\tilde{P}^{\mu\nu\alpha\beta}\nabla_\mu\nabla_\nu\phi\nabla_\alpha\nabla_\beta\phi
 =0
 \label{eq:dbi_unbraided}
\end{equation}
with
\begin{eqnarray}
\tilde L^{\mu\nu}&=&\mathcal G_0g^{\mu\nu} 
+\mathcal S_0 \phi^\mu\phi^\nu+\mathcal C_T T^{\mu\nu} \,, \label{eq:4_unbraided_open_indices_dbi}
\\ 
\tilde P^{\mu\nu\alpha\beta}&=&\mathcal V_{4B}\left(g^{\mu\nu}g^{\alpha\beta}-g^{\mu\alpha}g^{\nu\beta}\right)+\mathcal V_{4D}\left(\phi^\mu\phi^\alpha g^{\nu\beta}-\phi^\mu\phi^\nu g^{\alpha\beta}\right)\,,
 \label{eq:4_unbraided_higher_power_DBI}
\end{eqnarray}
where again the exact form of the coefficients can be found in the appendix \ref{sec:DBI_Debraided}.
Here we note some interesting features in the debraided field equation. First, the linear term contains only one coupling between $T^{\mu\nu}$ and the scalar field second derivative, while two terms couple to the trace of the energy-momentum tensor and its contraction with first derivatives of the field (as an effective potential). Second, it only contains non-linear derivative interaction terms of order $(\nabla\nabla\phi)^2$: the $(\nabla\nabla\phi)^3$ terms characteristic of quartic theories cancel in the debraiding procedure.
Finally, both the Weyl coupling and the non-linear second time derivatives cancel from the equations. Hence, the spurious solutions analysed in section \ref{sec:spurious_sol} are absent.%
\footnote{DBI theories have other interesting properties. For example, when regarded as effective theories they only require that the second and higher derivatives remain small $\phi^{(n)}\ll \Lambda^{n+1}$ while they allow for arbitrary values of first derivatives \cite{Burgess:2014lwa}.}

The reason for this simplicity is that the theory can be casted in a much simpler form by means of a field redefinition, as the DBI Galileon is equivalent to Einstein gravity plus a disformal coupling to matter \cite{Zumalacarregui:2012us,Bettoni:2013diz}. Because of this simplicity, DBI Galileons offer a toy example of kinetic mixing in more complicated Horndeski theories which also offers a viable and interesting alternatives to inflation \cite{Kaloper:2003yf} and as a mechanism for present day acceleration \cite{Zumalacarregui:2012us,Koivisto:2013fta,Sakstein:2014aca,vandeBruck:2015ida}.

If we further restrict the class of models and consider a constant $A$ in the definition of $G_4$ so that 
\begin{equation}
G_4=\frac{M_p^2}{2}\sqrt{1-2 \frac{X}{\Lambda^4}}\,,
\label{eq:dbi_G4_Kin}
\end{equation}
then the equation takes the even simpler form
\begin{equation}
\tilde{L}^{\mu\nu}\phi_{;\mu\nu} +\tilde{V}=0\,,
\label{eq:dbi_unbraided_Kin}
\end{equation}
where
\begin{eqnarray}
\nonumber
\tilde{L}^{\mu\nu}&=&\left(G_{2,X}+\frac{1}{\Lambda^4}\frac{G_2}{1-2X/\Lambda^4}\right) g^{\mu\nu}-\left(G_{2,XX}-\frac{1}{\Lambda^4}\frac{G_{2,X}}{1-2X/\Lambda^4}\right) \phi^{,\mu}\phi^{,\nu}
\nonumber \\ && +\frac{1}{\Lambda^4} \frac{1}{1-2X/\Lambda^4}T^{\mu\nu}\,,
\label{eq:dbi_Kin_L}
\\
\tilde{V} &=& G_{2,\phi}-2 X G_{2,\phi X}
\end{eqnarray}
where only the coupling between matter and second derivatives of the field remains.
Also notice that the non-linear derivative interaction terms have completelly vanished in this case. Thus this model does not have a Vainshtein like screening mechanism, althought it might possess a D-BIonic screening \cite{Burrage:2014uwa}. This model is equivalent to the simplest disformally coupled theories.

\section{Consequences of kinetic mixing} 
\label{COKM}

In this section we explore some of the consequences of kinetic mixing, namely the coupling to the Weyl tensor, the stability in the pressence of matter and the possibility of screening modifications of gravity by kinetic mixing.
The discussion will present both results for general quartic theories and their simpler DBI-like counterparts.

\subsection{Coupling to the Weyl tensor and existence of an Einstein frame} \label{sec:weyl}

One of the most salient features of quartic Horndeski theories, which is highlighted by our procedure, is the coupling between the Weyl tensor and the derivatives of the scalar field. The specific form of the coupling is
\begin{equation}\label{eq:weyl_coupling}
 \phi^{,\alpha}\phi^{,\beta} W^\mu{}_\alpha{}^\nu{}_\beta\phi_{;\mu\nu} = \mathcal E^{\mu\nu}\tilde\phi_{;\mu\nu}\,,
\end{equation}
where the right hand side identifies the electric part of the Weyl tensor with respect to the constant $\phi$ hypersurfaces.%
\footnote{This implicitly assumes that $\phi_{,\mu}$ is time-like. We thank I. Sawicki for pointing out the electric character of the coupling.}
Note that due to the symmetries of the Riemann tensor the above is the only independent contraction that one can form with derivatives of the scalar. By the traceless property of the Weyl tensor, it only couples to $\tilde\phi_{;\alpha\beta}=\phi_{;\alpha\beta}-\Box\phi g_{\alpha\beta}$ and $\tilde X_{\mu\nu} = \phi_{,\mu}\phi_{,\nu} +2Xg_{\mu\nu}$.

The Weyl tensor is not determined algebraically from the metric equations. Instead, it obeys a propagation equation
\begin{equation}
\nabla^\alpha W_{\mu\nu\alpha\beta}=\nabla_{[\mu}G_{\nu]\beta}+\frac{1}{3}g_{\beta[\mu}\nabla_{\nu]}G^\alpha{}_\alpha \label{eq:Weyl_propagation}\,,
\end{equation}
and it is therefore determined by the other fields through differential identities. One can generally decompose the curvature into a trace part (Ricci) and a traceless part (Weyl) using equation (\ref{eq:weyl_def}). In Einstein's theory, the Ricci curvature vanishes in the absence of matter. Therefore the Weyl tensor fully characterizes vacuum effects such as gravitational waves and tidal forces, both of them sourced by derivatives of the energy-momentum tensor (as given by the above propagation equation after substituting $G_{\mu\nu}\to 8\pi G T_{\mu\nu}$). One can formally invert eq. (\ref{eq:Weyl_propagation}) and write the Weyl tensor with a non-local dependence on the energy-momentum tensor.

In quartic theories, however, the direct coupling to the Weyl tensor implies that the scalar field has a new interaction with space-time curvature. The effects of this coupling are more difficult to interpret, since in ST theories non-trivial configurations of the scalar field can also produce non-zero Ricci curvature in the absence of matter. This feature prevents the Weyl tensor from fully characterizing curvature in the absence of matter, making it harder to link the coupling (\ref{eq:weyl_coupling}) to vacuum effects such as tidal forces or gravity waves.
However, in some situations the scalar field contribution to the metric equations might be subdominant with respect to matter, and we can recover the usual notion of the Weyl tensor describing such effects. At least in those cases, the Weyl coupling will affect the way in which the scalar field is sourced by the gravitational field produced by distant matter.

\begin{figure}[t]
\includegraphics[width=0.4\columnwidth]{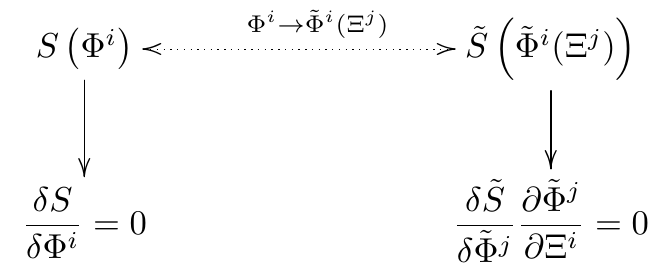}
\caption{Field redefinitions and linear transformations of the equations of motion.}\label{fig:field_redefinitions}
\end{figure}

Another important consequence of the occurrence of the Weyl tensor is that it obstructs the debraiding procedure, as it can not be obtained through contractions of the metric equations. This fact can be related to an important property of these kind of theories which is at the foundation of our classification into shaken and stirred theories.
The presence of the Weyl tensor in the equations of motion can be related to the lack of existence of a (local) field redefinition which renders the kinetic term for gravity canonical, i.e. of the Einstein--Hilbert form. The diagram in figure \ref{fig:field_redefinitions} shows how field redefinitions in the action correspond to linear transformation of the equations of motion, where the transformation matrix is given by the Jacobian of the field redefinition $\frac{\partial \tilde\Phi^j}{\partial\Xi^i}$ \cite{Zumalacarregui:2013pma}.
The Weyl tensor represents an element that can not be ``rotated away'', therefore indicating that the kinetic term for the metric (given by $\mc L_4$) can not be made canonical by a local field redefinition. Non-local redefinitions might be an exception, as they might ``undo'' the effects of the propagation equation for $W_{\mu\nu\alpha\beta}$ (\ref{eq:Weyl_propagation}).

Hence, the fact that DBI-like theories do not produce a coupling to the Weyl tensor is fully consistent with the existence of an Einstein frame for this class of theories. 
It is known that only a very special subclass of Horndeski theories can be related to Einstein gravity via a {local} field redefinition. In reference \cite{Bettoni:2013diz} this was investigated using general disformal transformation, showing that the most general quartic theory that can be mapped via a special disformal transformation to its Einstein frame version has to take the very special DBI-like form in which $G_4(\phi,X)=A(\phi)\sqrt{1-2B(\phi)X}$. An argument as to why more general field redefinitions involving the scalar field will not accomplish this goal is given in appendix \ref{sec:no_einstein_frame}.

\subsection{Stability in the presence of matter} \label{sec:stability}

An important application of the debraided field equation is to study the stability of the theory in the presence of matter.

When considering DBI Galileons, equation (\ref{eq:dbi_unbraided}) shows a flaw of the theory, as the kinetic mixing term might becomes problematic in the presence of large matter pressure. More precisely, if the energy-momentum tensor of matter contains a positive isotropic pressure term $T_{\mu\nu}\supset p g_{\mu\nu},\, p>0$, then the speed of sound of the field perturbations can become imaginary. For the simplest case (\ref{eq:dbi_G4_Kin}) with $G_2 = X$, the evolution equation (\ref{eq:dbi_unbraided_Kin}) reads
\begin{equation}\label{eq:dbi_pressure_unstable}
 \left(\Lambda^4-X\right)\Box\phi + \phi^{,\mu}\phi^{,\nu}\phi_{;\mu\nu} + T^{\mu\nu}\phi_{;\mu\nu}=0\,,
\end{equation}
where $1-2X/\Lambda^4\neq 0$ as it is related to the effective Planck mass (\ref{Eff_Planck_Mass}). One finds that the speed of sound squared (given by the coefficient of the second spatial derivatives) can become negative, leading to a gradient instability. The critical value of the pressure above which this happens is approximately
\begin{equation}
 p_c \sim \max(\Lambda^4, X)\,,
\end{equation}
where the estimate is obtained by looking at the sign of the $ii$ component of eq. (\ref{eq:dbi_pressure_unstable}), 
where it has been assumed that $\Lambda^4 >0$ in order to make the second time derivative coefficient positive when $\rho = T^{00}$ is large in order to avoid ghosts. The occurence of such instability is equivalent to the failure of the coefficient of $\phi_{;\mu\nu}$ in (\ref{eq:dbi_pressure_unstable}) to have Lorentzian signature and hence of the field equation to be hyperbolic.

DBI-like theories can therefore develop a gradient instability if the term proportional to the energy-momentum tensor dominates (see also \cite{Berezhiani:2013dw} for some remarks in this direction).
This problem is particularly acute if we are interested in theories able to explain cosmic acceleration, for which $\Lambda$ needs to be a very low energy scale. In this case we may spoil the predictions of homogeneous cosmology at early times, particularly during radiation domination when $p\sim \rho/3$. Therefore, the simplest solution is to raise the value of $\Lambda$ to make it higher than the reheating temperature.%
\footnote{Raising $\Lambda^4$ beyond the scale of reheating is not necessary: inflation requires a negative pressure and therefore leads to no gradient instability in equation (\ref{eq:dbi_pressure_unstable}). Ref. \cite{Kaloper:2003yf} considers an inflationary scenario in which $\Lambda^4$ is negative, but so large that the energy density never flips the sign of the kinetic term.}
Another solution is to give $\Lambda$ a field dependence (equivalent to the more general DBI-like theory (\ref{eq:dbi_unbraided})) so that the critical pressure is always larger than the cosmological one.%
\footnote{The instability might be resolved by self-consistently accounting for the evolution of matter. However, even if the dynamics does not lead to singularities, it may cause large inhomogeneities incompatible with cosmological observations (see \cite{Amendola:2015tua} for an example).}

The last possibility is the re-introduction of non-linear derivative self-interactions by modifying the cubic term such that $G_{3,X}\neq 0$, without modifying $G_4$. These terms would dominate when the spatial derivatives become large, and may act to stabilize the equation at some finite wave number, given by the condition
\begin{equation}
 p/\Lambda^4 \sim k^2 G_{3,X}\delta\phi\,.
\end{equation}
This estimate relies on te scaling of the derivative self-interactions in a cubic theory, which are $\sim G_{3,X} (\nabla\nabla \phi)^2\propto k^4$. Even if this modification stabilizes the perturbations, it would introduce large spatial gradients which might spoil the homogeneity and affect cosmological observables.

Finally, it is possible to avoid the gradient instability by making a different choice of $G_4$. As it was shown in section \ref{section:quartic_mixing}, theories different from the DBI Galileon have a richer mixing structure, as given by equation (\ref{eq:quartic_debraided}). The additional terms present in the general case may balance the ones leading to the gradient instability. For example, the general debraided equations for a quartic theory contain a term $\mc G_T T\Box\phi$, for which the spatial and time derivatives have always the correct relative sign and no gradient instabilities occur. It is therefore possible to avoid the gradient instabilities by constructing a theory in which such terms are sufficiently large. 
Of course, in order to assess the viability of these models, a full dynamical analysis is required, but this goes beyond the scope of the present work and is left for further studies.

\subsection{Screening scalar forces by kinetic mixing}\label{section:screening}

Alternative theories of gravity typically introduce additional forces, which may alter the predictions in the local system and render them incompatible with local gravity tests (for a review, see ref. \cite{Joyce:2014kja}). However, some theories provide \emph{screening mechanism}, which hide the effects of scalar forces via non-linear interactions of the field. Although these mechanisms have been mostly studied by making these couplings explicit (e.g. working in the Einstein frame), they can also be identified in a minimally coupled description \cite{Hui:2009kc}.

We conclude the discussion of the consequences of kinetic mixing by noting that the structure of the mixing terms allows to generalize the previously proposed \emph{disformal screening mechanism} to a more general phenomenon, based on kinetic mixing properties and present in a larger class of theories.
The disformal screening mechanism was introduced in the context of disformally coupled theories, which are the Einstein-frame version of (and therefore equivalent to) the DBI-like Galileons that we considered in section \ref{sec:DBI_galileon}. Its action is based on two observations, see eq (\ref{eq:dbi_unbraided})
\begin{enumerate}
 \item If the field is static (no time derivatives) and only disformally coupled ($M_p$ is constant) then the field decouples from non-relativistic matter \cite{Noller:2012sv}.
 \item If the field evolves in time and the energy density is non-relativistic (only $T^{00}= \rho$ contributes significantly), the field evolution becomes independent of the energy density (see refs. \cite{Koivisto:2012za,Zumalacarregui:2012us} for details). This property does not rely on the specific form of the conformal and the disformal coupling.
\end{enumerate}
The efficiency of this mechanism to reconcile DBI-like theories with local gravity tests is difficult to investigate in practice, as it requires considering simultaneously spatial and time dependence (demanding that the time-evolution is a sub-dominant effect clearly spoils the existence of a screened solution \cite{Sakstein:2014isa}, although numerical studies seem to confirm the screening effects \cite{hagala:2015paa}). Moreover, pure DBI-like theories have to face the issues related to stability in the presence of matter with non-negligible pressure, as described in section \ref{sec:stability}, as well as other difficulties to fulfil laboratory tests \cite{Brax:2014vva}.

Ultimately, the main requirement for the disformal screening mechanism is that the coefficients of $\phi_{;\mu\nu}$ depend on the energy-momentum tensor.
The identification terms containing the matter energy-momentum tensor in the second derivatives of the scalar in the debraided field equation (\ref{eq:debraided_L_term}) indicates that the disformal screening mechanism, or variations thereof, can occur in a much larger class of quartic Horndeski theories. Among others, those terms include contributions proportional to
\begin{equation}
 T^{\mu\nu}\phi_{\mu\nu},\; T\Box\phi,\; \phi^{,\alpha}T_{\alpha\beta}\phi^{,\beta}\Box\phi,\; T\phi^{,\alpha}\phi^{,\beta}\phi_{;\alpha\beta}\; \cdots\,,
\end{equation}
where only the first one is present in DBI-like theories and the disformal screening mechanism. Therefore, the richer braiding structure of quartic Horndeski theories has the potential to soften some of the requirements needed for the disformal screening and alleviate some of its problems. Because of its extended generalities beyond disformally coupled/DBI-like theories, we propose referring to this mechanism as \emph{screening by kinetic mixing}. 

\section{Beyond Quartic Theories} \label{section:beyond_quartic}

In this section we comment on the issue of kinetic mixing in more general ST theories. We will not go into the same level of details as for the quartic, limiting our analysis to the mixing terms and pointing out the novel structures that appear.

\subsection{Quintic Horndeski theories: Non-linear mixing} \label{section:quintic_theories}

In the case of quintic Horndeski theories we expect terms quadratic in the curvature to be present in the scalar field equation. Due to the second order nature of the theory we will only find terms that do not introduce higher derivatives of degrees of freedom. One example of this is the Gauss-Bonnet term $\mathcal G = R^2-4R_{\mu\nu}R^{\mu\nu} + R_{\mu\nu\alpha\beta}R^{\mu\nu\alpha\beta}$, that would be generated in the field equation as
\begin{equation}
 \mathcal L_{GB}\supset f(\phi) \mathcal G \to f' \mathcal G \delta\phi\,,
\end{equation}
since the above Lagrangian is equivalent to a particular choice of Horndeski functions with $G_{5,X}\neq 0$ \cite{Kobayashi:2011nu}.

These theories hence produce a new form of \emph{non-linear mixing}. Covariantly debraiding the field equation will therefore require (at the very least) contractions of the metric equations with the Riemman and the Ricci tensor, and then further contractions similar to those needed in section \ref{section:quartic_mixing}. Whether such terms can be debraided using the techniques introduced in the previous sections lies beyond the scope of this work, but at the very least we can anticipate the difficulties we already encountered for the quartic theories. In particular, the coupling to the Weyl tensor occurs already at the level of the action. This is not manifest from the simpler form shown in eq. (\ref{LH5}), but can be obtained in an equivalent form obtained by integrating by parts (e.g. in Horndeski's original paper \cite{Horndeski:1974wa}, where dual of the Riemann tensor appears in the action). 
 
\subsection{Theories beyond Horndeski: derivative mixing} \label{section:beyond_horndeski}

The previous sections have shown how Horndeski theories feature a form of kinetic mixing that is \emph{algebraic} in the energy-momentum-tensor. Healthy non-Horndeski theories \cite{Zumalacarregui:2013pma,Gleyzes:2014dya} display a novel form of kinetic mixing, in which the energy-momentum tensor enters the scalar field equation through its \emph{derivatives}. These theories have received attention recently, including studies in the context of late time acceleration \cite{Gleyzes:2014qga,DeFelice:2015isa}, inflation \cite{Fasiello:2014aqa} and local gravity tests \cite{Kobayashi:2014ida,Saito:2015fza,Koyama:2015oma}.

The simplest examples of theories beyond Horndeski are the ones originally proposed by Bekenstein \cite{Bekenstein:1992pj}.
These are ST theories with an Einstein--Hilbert term for gravity, a k-essence Lagrangian and a matter Lagrangian constructed out of a metric that explicitly involves the scalar field. 
\begin{equation}\label{eq:action_bekenstein}
 S_{\rm B}[\bar g_{\mu\nu},\phi,\psi]=\int d^4 x\left(\sqrt{-\bar g}\frac{M_p^2}{2} \bar R[\bar g_{\alpha\beta}] + \sqrt{-\bar g}G_2(\bar X,\phi) 
 + \sqrt{-\tilde g}\mathcal L_m(\tilde g_{\mu\nu},\psi )\right)\,,
\end{equation}
The novel ingredient is that the matter Lagrangian is constructed using a general disformal metric
\begin{equation}\label{eq:disformal-general}
 \tilde g_{\mu\nu}[\bar g_{\alpha\beta},\phi]=C(\bar X,\phi)\bar g_{\mu\nu} + D(\bar X,\phi)\phi_{,\mu}\phi_{,\nu}\,.
\end{equation}
When written in the Jordan frame via a non-trivial inversion of eq. (\ref{eq:disformal-general}), these theories have been shown to be non-equivalent to any Horndeski theory \cite{Bettoni:2013diz,Zumalacarregui:2013pma} unless $C_{,X},D_{,X}=0$ 
Indeed, their Euler-Lagrange variation yields equations with derivatives higher than second order.

This seems to suggest that the theory propagates an extra degree of freedom. However, it was found that an implicit constraint exists in the equations of motion for the metric which allows to remove all the higher time derivatives from the equations of motion and cast the dynamical equations in a second order form. This procedure uses a contraction of the metric equations in a manner analogous to the procedure performed in section \ref{section:quartic_mixing}, and thus the formulation using implicit constraints (which is second order) introduces the energy-momentum tensor in the equations of motion. Here we sketch the basics, the interested reader is referred to ref. \cite{Zumalacarregui:2013pma} and appendix \ref{sec:BH_examples} for further details.%
\footnote{See \cite{Wetterich:2014bma} for similar remarks in a more general setting and a generalization of Bekenstein's program to find the most general ghost-free theories that are related to Einstein gravity by a field redefinition. Some of the generalizations proposed there have also been explored in ref. \cite{Amendola:2010bk}.}

The novelty of this class of theories is that the terms introduced with this procedure involve derivatives of the energy-momentum tensor in the field equation, therefore providing a new form of kinetic mixing. This can be seen by considering $S_{\rm B}[\tilde g_{\mu\nu},\phi,\psi]$, the Jordan frame version of (\ref{eq:action_bekenstein}), in which the equations of motion are obtained for $\tilde g_{\alpha\beta}$ (rather than $\bar g_{\mu\nu}$) using the inverse of the matter metric (\ref{eq:disformal-general}). The inverse relation between the metric has the same structure, $\bar g_{\mu\nu}[\tilde g_{\alpha\beta},\phi] = A(\tilde X,\phi) \tilde g_{\mu\nu} + B(\tilde X,\phi) \phi_{,\mu}\phi_{,\nu}$, where  $A,B$ can be obtained implicitly given the form of $C,D$. We note that it is possible (although cumbersome) to express the Jordan frame theory in terms of $\tilde g_{\alpha\beta}$. This is done for two simple cases in appendix \ref{sec:BH_examples}.

By using the chain rule and the properties of the Jacobian of the transformation between the metrics, it is possible to write the field equation without higher order field derivatives as
\begin{equation}\label{eq:JF_gen_disf_2nd}
\bar \nabla_\al\lp \mc T_{\rm K} \tilde\phi^{,\al}\rp - \bar G^{\mu\nu}\bar \nabla_\mu\lp B\phi_{,\nu}\rp 
+ \bar G^{\mu\nu}\lp A_{,\phi} g_{\mu\nu} + B_{,\phi}\phi_{,\nu}\phi_{,\mu}\rp 
- \sqrt{\frac{\tilde g}{\bar g}}\frac{\delta \Lag_\phi}{\delta \phi} = 0\,,
\end{equation}
where $A,B$ define the inverse of (\ref{eq:disformal-general}) and all barred quantities are meant to be evaluated in terms of $\tilde g_{\mu\nu}$ using this relation. The \emph{kinetic mixing factor} is defined as
\begin{equation}\label{eq:kinmix_JF_gen}
 \mc T_{\rm K} \equiv \frac{1}{M_p^2}\sqrt{\frac{\tilde g}{\bar g}}
 \frac{ \big(A_{,\tilde X} \tilde g_{\mu\nu} + {B_{,\tilde X}}\phi_{,\mu}\phi_{,\nu} \big)\big(\tilde T^{\mu\nu}_{\rm m}+\tilde T^{\mu\nu}_{G_2}\big)}{ A - A_{,\tilde X}\tilde X + 2 B_{,\tilde X} \tilde X^2}\,,
\end{equation}
where the energy-momentum tensor has the usual definition in terms of a variational derivative with respect to $\tilde g_{\mu\nu}$. This  factor enters through the first term in the field equation (\ref{eq:JF_gen_disf_2nd}) and thus introduces a derivative form of mixing between the matter and the field, even though the scalar and matter fields are minimally coupled. Of course, it remains to be proved that the curvature stemming from the barred Einstein tensor $\bar G^{\mu\nu}$ can be traded for algebraic couplings to matter and the scalar field in an analogous way as it has been done for the quartic Galileon. In appendix \ref{sec:BH_pure_conformal} it is shown how for the pure conformal coupling this can be done, thus achieving a second order fully debraided equation for a beyond-Horndeski model (see ref. \cite{Motohashi:2015pra} for an explicit debraided form in a more general case).

We can think of the Jordan Frame version of Bekenstein's theories as a simple playground to study the features of theories beyond Horndeski. In this sense they are analogous to the simple DBI Galileons discussed in section \ref{sec:DBI_galileon}. More general theories beyond Horndeski will not accept a simple Einstein frame formulation and will therefore not be as simple to unbraid. We expect that generalizations of Bekenstein theories will contain a richer mixing structure reflecting their more diverse phenomenology.
Other theories beyond Horndeski have been proposed by Gleyzes, Langlois, Piazza and Vernizzi \cite{Gleyzes:2014dya}, where it is also shown that no additional degrees of freedom are introduced and for which similar findings regarding kinetic mixing have been reported \cite{Gleyzes:2014dya,Gleyzes:2014qga}. In particular, the authors found terms describing the interactions between the scalar and derivatives of the matter energy density on perturbed cosmological backgrounds. See also refs. \cite{Gao:2014soa,Gao:2014fra} for further work on extensions beyond Horndeski.

\section{Conclusions and outlook}
\label{CAO}
In an era in which the alternatives to Einstein gravity are getting more and more complex, with an increasing level of mixing between the various degrees of freedom that build the theory, it is of fundamental importance to develop methods that allow the classification and facilitate the study of different models. This will not only lead to a better understanding of the physical content of a theory, but also can shed light on its properties, potential issues and observable consequences.

In this work we have considered a method, covariant debraiding, to study the kinetic mixing between the scalar and tensor degrees of freedom in general, alternative theories of gravity. Our method consists in using contractions of the metric equations of motion to remove Ricci-curvature terms in the field equation of motion. This approach relies on the full equations of motion, and therefore allows to draw conclusions regardless of any approximation scheme, in a fully non-linear fashion with the extra advantage of being in the Jordan frame where the energy-momentum tensor of matter is covariantly conserved. Hence, the debraiding procedure provides a useful way to study the properties of ST theories and their interaction with matter as well as the stability of the scalar field rather directly.

As an application of the method, we have extended covariant debraiding for the first time beyond the simplest examples and applied it to the study of quartic Horndeski theories in detail. 
These theories display a new set of mixing terms, which indicate new forms of coupling of the scalar field. The novel terms appearing in quartic Horndeski theories involve the contraction of second derivatives of the scalar with curvature which translate into contractions of $\phi_{;\mu\nu}$ and $T_{\mu\nu}$ in the debraided equation. The procedure allows one to study how matter sources the scalar field despite both being coupled minimally, interacting only with the metric directly. 

General quartic theories also feature a coupling to the curvature that can not be removed by covariant debraiding. This term is given by the Weyl tensor, which is not algebraicly determined from the metric equations, contracted with the first and second derivatives of the scalar. It represents a novel form of interaction between the scalar field and space-time in the absence of matter. 
This form of coupling to the vacuum and the non-local nature of the Weyl tensor{, determined by the global distribution of matter, might have relevant implications for Mach's principle in ST theories, to be addressed in a follow up work.
In addition, the debraided equation generally contain non-linear derivative terms without antisymmetric structure. These may introduce spurious solutions to the equations of motion, which can be nonetheless distinguished from the physical ones and disregarded. However, this problem is absent on sufficiently simple situations, such as  cosmological backgrounds.

Covariant debraiding singles out a particular subset of theories for which the debraiding equation neither contains the Weyl coupling, nor leads to spurious solutions. These theories generalize the quartic DBI Galileon, and coincide with the maximal set of Horndeski theories which accept an Einstein frame formulation (i.e.  the kinetic term for the tensor degree of freedom can be written in the Einstein Hilbert form). This relation to Einstein gravity is behind the lack of Weyl coupling: as field redefinitions at the level of the action are equivalent to linear transformations of the equations of motion, the occurrence of a term that can not be ``rotated away'' indicates the non-existence of a transformation to a non-mixed frame.

 Although the equations are far too involved for an analysis of general quartic theories, interesting conclusions can be easily drawn for the simple DBI-like models. Using the debraided equations we show that quartic DBI Galileons have gradient instabilities in the presence of matter with sufficiently large pressure. This would spoil the early universe predictions unless the energy scale that suppresses the coupling is very large or the theory is extended beyond the simplest case. In particular, a non-DBI quartic theory has a richer mixing structure, including terms which can stabilize the gradient instabilities. 
The debraided form of the equations that we present here can be used to design models with certain properties, by choosing the Horndeski functions to enhance a particular set of terms.

Other features of quartic Horndeski theories are further clarified using their debraided formulation. One example is the screening of scalar forces by kinetic mixing. This mechanism was first investigated for DBI-like theories in the Einstein frame, and known as disformal screening mechanism. Our work implies that this effect is not exclussive of DBI-like/disformally coupled theories, but rather ubiquitous in quartic Horndeski theories. Moreover, more general theories might weaken the assumptions necessary for the screening mechanism to be effective.

The mixing terms in the equations of motion are characteristic of each theory and carry the information about how the scalar degree of freedom interacts with matter. These terms grow in complexity in quartic theories (including the Weyl tensor coupling) and beyond, leading to non-linear mixing in quintic theories and derivative mixing in theories beyond Horndeski. 
The covariant debraiding procedure also provides a binary classification of models according to their kinetic mixing properties: Theories for which the covariant debraiding eliminates all instances of the curvature are \emph{stirred}. This includes old-school theories, cubic and DBI-like quartic theories and some simple non-Horndeski theories. More generally, theories in which some residual curvature remains after covariant debraiding are \emph{shaken}, including general quartic theories (featuring the Weyl tensor), quintic theories and general theories beyond Horndeski.

There are other, potentially interesting avenues for further development. Our work has focused only on a very restricted form of debraiding, one in which locality and Lorentz invariance are manifest. One can use more general procedures to debraid the equations, e.g. by choosing a particular time slicing and solving for the highest time derivatives. This can be achieved (by explicit choice of coordinates or by an ADM decomposition), and is indeed necessary if one aims to numerically solve the general equations. As a step beyond, one may consider \emph{non-local} debraiding by formally solving for Weyl tensor in terms of its propagation equation. These and other developments might  shed light on a number of problems, such as the initial value formulation of modified gravity theories.

The main lesson to be learned is that gravitational degrees of freedom become more fundamentally mixed the further away we go from Einstein's theory of gravity. Any departure from the canonical kinetic term for the metric tensor necessarily introduces at least a scalar degree of freedom, which can then interact in a variety of ways, ranging from Brans-Dicke to beyond Horndeski.
For both stirred to shaken theories, this increasing complexity reflects on the fundamental Lagrangian and gives important hints about its nature and dynamics.
The study of kinetic mixing provides new ways to classify models and address their properties, valid across generations of ST theories, and will provide a useful tool in the understanding of alternative gravities.

\acknowledgments{
 The authors wish to thank I. Sawicki and J. Beltran for their careful reading of a previous version of this draft and L. Amendola, E. Bellini, J. Gleyzes, T. Koivisto, J. Sakstein, A. Solomon, C. Wetterich for all the useful discussions and suggestions. The authors thank Nordita for hosting them during an important part of the collaboration. The computer algebra package xAct \cite{Brizuela:2008ra,xAct} has been used in some of the computations. MZ acknowledges support from TRR33 The Dark Universe. DB acknowledges support from the I-CORE Program of the Planning and Budgeting Committee, THE ISRAEL SCIENCE FOUNDATION (grants No. 1829/12 and No. 203/09), and the Asher Space Research Institute.}

\appendix

\section{Quartic Horndeski theories}
\label{section:full_eqs_quartic}
In this appendix we present the general equations for a Quartic Horndeski action. This kind of action assumes $G_3(\phi,X)=0=G_5(\phi,X)$ while leaves the other two functions generic. We will first write the equations as derived from the variation of the Quartic Horndeski action (\ref{LH4}) and after present the detailed structure of the coefficients of the debraided equations that appears in (\ref{eq:unbraided_schematic}).

\subsection{Equations for quartic theory}
In this section we derive the both the metric and the scalar field equations in their explicit form. To simplify the expressions we will use the following notation for contractions $[\phi^n]=g^{\mu\nu}\phi^n_{\mu\nu}$, $\enangle{\phi^n}=\phi^{,\mu} \phi^n_{\mu\nu}\phi^{,\nu}$, $\phi^n_{\mu\nu}=\phi_{;\mu\alpha_1}\phi\ud{;\alpha_1}{;\alpha_2}\cdots\phi\ud{;\alpha_{n-1}}{;\nu}$, $\enangle{R}=\phi^{,\alpha}\phi^{,\beta}R_{\alpha\beta}$ (if $F_{\alpha\beta}$ has two indices), $\enangle{W}=\phi^{,\alpha}\phi^{,\beta}W_{\alpha\mu\beta\nu}\phi^{;\mu\nu}$, $\enangle{R\phi}=\phi^{,\alpha}R_{\alpha\lambda}\phi^{;\lambda\beta}\phi_{,\beta}$.

The variation of the Lagrangians (\ref{LH2}) and (\ref{LH4}) with respect to the scalar field, gives the following equation for the scalar field

\begin{eqnarray}
\nonumber
G_{2,\phi}&-&2G_{2,\phi X}X + \left(G_{2,X}-4G_{4,\phi\phi X}X\right)\Box\phi-\left(G_{2,XX}+2G_{4,\phi\phi X}\right)\phi^{,\mu}\phi^{,\nu}\phi_{;\mu\nu}\\
\nonumber
&+&2G_{4,\phi XX}\left(2\enangle{\phi^2}-2\enangle{\phi}[\phi]-\left([\phi]^2-[\phi^2]\right)X\right)+G_{4,XX}\left([\phi]^3-3[\phi][\phi^2]+2[\phi^3]\right)\\
\nonumber
&+&G_{4,XXX}\left(-2\enangle{\phi^3}+2\enangle{\phi^2}[\phi]-\enangle{\phi}\left([\phi]^2-[\phi^2]\right)\right)+3G_{4,\phi X}\left([\phi]^2-[\phi^2]\right)\\
\nonumber
&+&G_{4,\phi}R -2G_{4,X}G^{\mu\nu}\phi_{;\mu\nu}-2G_{4,\phi X}\left(2\enangle{R}+RX\right)\\
\nonumber
&+&G_{4,XX}\left(-[\phi]\enangle{R}-\frac{2}{3}R\left(\enangle{\phi}-[\phi]X\right)+2\enangle{W}-2\enangle{R\phi}X+2\phi^\alpha\phi^\beta R_{\beta\gamma}\phi_\alpha{}^\gamma\right)=0\,.\\
\end{eqnarray}

The variation with respect to the metric gives
\begin{eqnarray}
\nonumber
&& G_4 G_{\alpha\beta}-\frac{1}{2}T_{\alpha\beta}^{(m)}+\left(-\frac{G_{2,X}}{2}-G_{4,\phi\phi}-2G_{4,\phi X}[\phi]-\frac{1}{2}\left([\phi]^2-[\phi^2]\right)-\frac{1}{3}G_{4,X}R\right)\phi_\alpha\phi_\beta\\
\nonumber
&+&\left(-\frac{1}{2}G_2+G_{4,\phi}[\phi]+\frac{1}{2}G_{4,X}\left([\phi]^2-[\phi^2]-\enangle{R}\right)-2G_{4,\phi\phi}X+\frac{1}{3}RG_{4,X}X\right.\\
\nonumber
 &&\left. -2G_{4,\phi X}\left(\enangle{\phi}+[\phi]X\right)+G_{4,XX}\left(-\enangle{\phi}[\phi]+\enangle{\phi^2}\right)\right)g_{\alpha\beta}\\
 \nonumber
 &+&\phi_ {\alpha\beta}\left(-G_{4,\phi}+G_{4,XX}\enangle{\phi}-G_{4,X}[\phi]+2G_{4,\phi X}X\right)\\
 \nonumber
 &+&2G_{4,\phi X}\left(\phi^\gamma \phi_{\gamma\beta}\phi_\alpha+\phi^\gamma \phi_{\gamma\alpha}\phi_\beta\right)+G_{4,X}\left(\phi_\alpha{}^\gamma\phi_{\gamma\beta}-R_{\alpha\beta}X+\frac{1}{2}R_{\beta\gamma}\phi^\gamma\phi_\alpha+\frac{1}{2}R_{\alpha\gamma}\phi^\gamma\phi_\beta+W_{\alpha\mu\beta\nu}\phi^\mu\phi^\nu\right)\\
 &-&G_{4,XX}\left(\phi^\gamma\phi_{\beta}{}^\eta\phi_\gamma{}_\eta\phi_\alpha+\phi^\gamma\phi_{\alpha}{}^\eta\phi_\gamma{}_\eta\phi_\beta-\phi^\gamma(\phi_{\gamma\beta}\phi_\alpha+\phi_{\gamma\alpha}\phi_\beta)[\phi]+\phi^\gamma\phi_{\gamma\alpha}\phi_{\beta\eta}\phi^\eta\right)=0\,.
\end{eqnarray}

\subsection{Coefficients in the debraided equations}\label{sec:debraided_quartic_coefficients}

In this appendix we report the explicit expressions for the coefficients of debraided equation (\ref{eq:unbraided_schematic}). These coefficients will be fixed once a particular model is chosen, i.e., a choice for the form of $G_4$ is made.

The purely field dependent coefficients of the linear second order derivative are
\begin{eqnarray}
 \mathcal G_0 &=&\frac{1}{3}(G_{4} - 2 G_{4,X} X)^{-2} 
 \Big( G_{2,X} \bigl(3 G_4^2 + 8 G_{4,X}^2 X^2 - 4 G_4 X (3 G_{4,X} + G_{4,XX} X)\bigr) + G_2 \bigl(4 G_{4,X} X (2 G_{4,X} \nonumber \\ 
&& + 3 G_{4,XX} X) -  G_4 (3 G_{4,X} + 4 G_{4,XX} X)\bigr) - 3 \Bigl(4 G_4^2 G_{4,\phi \phi X} X + G_4 \bigl(-3 G_{4,\phi}^2 \nonumber \\ 
&& + 4 G_{4,\phi} G_{4,\phi X} X + 4 X (G_{4,X} G_{4,\phi \phi} + 5 G_{4,\phi X}^2 X + 2 G_{4,XX} G_{4,\phi \phi} X - 4 G_{4,X} G_{4\phi \phi X} X)\bigr) \nonumber \\ 
&& + 8 G_{4,X} X \bigl(G_{4\phi}^2 + X (- G_{4,X} G_{4,\phi \phi} - 4 G_{4\phi X}^2 X - 2 G_{4,XX} G_{4\phi \phi} X \nonumber \\ 
&& + 2 G_{4,X} G_{4,\phi \phi X} X)\bigr)\Bigr)\Big)
\,,
 \\ 
 \mathcal G_T &=&
 - \frac{(G_{4,X}^2 + G_{4} G_{4,XX}) X}{3 (G_{4} - 2 G_{4,X} X) (G_{4} -  G_{4,X} X)}
 \,,
 \\
 \mathcal G_{\enangle{T}} &=& 
 \frac{(G_{4,X}^2 + G_{4} G_{4,XX}) (-3 G_{4} + 4 G_{4,X} X)}{6 (G_{4} - 2 G_{4,X} X)^2 (G_{4} -  G_{4,X} X)}
 \,.
 \end{eqnarray} 
\begin{eqnarray}
 \mathcal S_0 &=&\frac{1}{3}\Bigl(-3 G_{2,XX} (G_4 - 2 G_{4,X} X)^2 - 6 G_{4,\phi \phi X} (G_4 - 2 G_{4,X} X)^2 + 6 G_{4,X} G_{4,\phi \phi} (- G_4 + 2 G_{4,X} X) \nonumber \\ 
&& + 12 G_{4,\phi X}^2 X (-4 G_4 + 7 G_{4,X} X) + G_{4,X} (-3 G_{2,X} G_4 + G_2 G_{4,X} - 3 G_{4,\phi}^2 + 4 G_{2,X} G_{4,X} X) \nonumber \\ 
&& + G_{4,\phi X} (-24 G_4 G_{4,\phi} + 36 G_{4,X} G_{4,\phi} X) + G_{4,XX} \bigl(G_2 G_4 + 12 G_{4\phi \phi} X (- G_4 + 2 G_{4,X} X) \nonumber \\ 
&& + 4 G_{2,X} X (-2 G_4 + 3 G_{4,X} X)\bigr)\Bigr)(G_{4} - 2 G_{4,X} X)^{-2}
\,,
 \\ 
 \mathcal S_T &=& 
 - \frac{G_{4,X}^2 + G_{4} G_{4,XX}}{6 (G_{4} - 2 G_{4,X} X) (G_{4} -  G_{4,X} X)}
 \,,
 \\
 \mathcal S_{\enangle{T}} &=& 
 - \frac{G_{4,X}^3 + G_{4} G_{4,X} G_{4,XX}}{6 (G_{4} - 2 G_{4,X} X)^2 (G_{4} -  G_{4,X} X)}
 \,.
\end{eqnarray}

The other kinetic couplings to matter/curvature read:
\begin{eqnarray}
 \mathcal C_T &=& 
 - \frac{G_{4,X} + G_{4,XX} X}{G_4 -  G_{4,X} X}
 \,, \\
 \mathcal C_{\enangle{T}} &=& 
 \frac{G_{4,X}^2 + G_4 G_{4,XX}}{(G_4 - 2 G_{4,X} X) (G_4 -  G_{4,X} X)}
 \,. \\
 \mathcal C_W &=& 
 \frac{2 (G_{4,X}^2 + G_4 G_{4,XX})}{G_4 -  G_{4,X} X}
 \,,
 \end{eqnarray}
 
Finally, there will be the generalizations of the (non-kinetic) conformal and disformal coupling, that we observed in KGB theories.
\begin{eqnarray}
 \mathcal Q_T &=& - \frac{G_{4,\phi} + 2 G_{4,X\phi} X}{2( G_4 - 2 G_{4,X} X)} \,, \\
 \mathcal Q_{\enangle{T}} &=& - \frac{4 G_4 G_{4,X\phi} + G_{4,X} (G_{4,\phi} - 6 G_{4,X\phi} X)}{2 (G_4 - 2 G_{4,X} X)^2}\,,
\end{eqnarray}
The potential term is
\begin{eqnarray}
\tilde V&=&\Bigl(G_{2,\phi} (G_4 - 2 G_{4,X} X)^2 + G_2 \bigl(-2 G_{4} G_{4,\phi} + G_{4,X} X (5 G_{4,\phi} + 2 G_{4,\phi X} X)\bigr) + X \bigl(-2 G_{2,\phi X} (G_4 \nonumber \\ 
&& - 2 G_{4,X} X)^2 - 6 G_{4,\phi \phi} (G_4 - 2 G_{4,X} X) (G_{4,\phi} + 2 G_{4,\phi X} X) + G_{2,X} (G_4 G_{4\phi} - 4 G_{4,X} G_{4,\phi} X \nonumber \\ 
&& - 6 G_{4} G_{4,\phi X} X + 8 G_{4,X} G_{4,\phi X} X^2)\bigr)\Bigr)(G_4 - 2 G_{4,X} X)^{-2}\,,
\end{eqnarray}
The coefficients in front of the non-linear derivatives are
\begin{eqnarray}
\nonumber
\mathcal{V}_{B4}&=&\Bigl(G_4^3 (3 G_{4,\phi X} - 2 G_{4,\phi XX} X) + G_4 G_{4,X} X \bigl(G_{4,XX} X (-13 G_{4,\phi} + 30 G_{4,\phi X} X) \\ \nonumber
&-& 4 G_{4,X} (3 G_{4,\phi} - 11 G_{4,\phi X} X + 4 G_{4,\phi XX} X^2)\bigr) + G_{4,X}^2 X^2 \bigl(12 G_{4,XX} X (G_{4,\phi} - 2 G_{4,\phi X} X) \\ \nonumber
&+& G_{4,X} (11 G_{4,\phi} - 30 G_{4,\phi X} X + 8 G_{4,\phi XX} X^2)\bigr) + G_{4}^2 \bigl(G_{4,XX} X (3 G_{4,\phi} - 10 G_{4,\phi X} X) \\
&+& G_{4,X} (3 G_{4,\phi} - 21 G_{4,\phi X} X + 10 G_{4,\phi XX} X^2)\bigr)\Bigr)(G_4 - 2 G_{4,X} X)^{-2} (G_4 -  G_{4,X} X)^{-1}\,,
\end{eqnarray}
\begin{eqnarray}
\mathcal{V}_{4D}&=&\Bigl(4 G_4^3 G_{4,\phi XX} + G_4^2 (3 G_{4,XX} G_{4,\phi} + 12 G_{4,X} G_{4,\phi X} + 22 G_{4,XX} G_{4,\phi X} X \nonumber \\ 
&& - 20 G_{4,X} G_{4,\phi XX} X) + G_{4,X}^2 X \bigl(48 G_{4,XX} G_{4,\phi X} X^2 + G_{4,X} (-5 G_{4,\phi} + 30 G_{4,\phi X} X \nonumber \\ 
&& - 16 G_{4,\phi XX} X^2)\bigr) + G_4 G_{4,X} \bigl(- G_{4,XX} X (5 G_{4,\phi} + 66 G_{4,\phi X} X) + G_{4X} (3 G_{4,\phi} \nonumber \\ 
&& - 38 G_{4\phi X} X + 32 G_{4,\phi XX} X^2)\bigr)\Bigr)(G_4 - 2 G_{4,X} X)^{-2} (G_4 -  G_{4,X} X)^{-1}\,,
\end{eqnarray}
\begin{eqnarray}
\mathcal{V}_{B5}&=&\frac{G_{4,X}^2+G_4 G_{4,XX}}{G_4-G_{4,X}X}\,,\\
\mathcal{V}_{D5}&=& \Bigl(3 G_4^3 (G_{4,\phi X} - 2 G_{4\phi XX} X) + G_4 G_{4,X} X \bigl(8 G_{4XX} X (- G_{4,\phi} + 12 G_{4,\phi X} X) \nonumber \\ 
&& + G_{4,X} (-15 G_{4,\phi} + 82 G_{4,\phi X} X - 48 G_{4,\phi XX} X^2)\bigr) + 4 G_{4,X}^2 X^2 \bigl(3 G_{4,XX} X (G_{4,\phi} \nonumber \\ 
&& - 6 G_{4,\phi X} X) + G_{4,X} (4 G_{4,\phi} - 15 G_{4,\phi X} X + 6 G_{4,\phi XX} X^2)\bigr) + G_{4}^2 \bigl(-32 G_{4,XX} G_{4,\phi X} X^2 \nonumber \\ 
&& + 3 G_{4,X} (G_{4,\phi} - 11 G_{4,\phi X} X + 10 G_{4,\phi XX} X^2)\bigr)\Bigr)(G_4 - 2 G_{4,X} X)^{-2} (G_4 -  G_{4,X} X)^{-1} \,,\nonumber \\
\end{eqnarray}

The remaining terms are those we have labeled as ``worrying terms'', as they might lead to spurious solutions on certain backgrounds (cf. \ref{sec:spurious_sol}). They read
\begin{equation}
\mathcal{W}_1 = \frac{(G_{4,X}^2 + G_4 G_{4,XX}) (G_{4,X} + 2 G_{4,XX} X)}{3 (G_4 - 2 G_{4,X} X) (G_4 -  G_{4,X} X)}\,,
\end{equation}
\begin{equation}
\mathcal{W}_2=\frac{(G_{4,X}^2 + G_{4} G_{4,XX})^2}{3 (G_4 - 2 G_{4,X} X)^2 (G_4 -  G_{4,X} X)}\,,
\end{equation}
\begin{equation}
\mathcal{W}_{D2}=- \frac{(G_{4,X}^2 + G_4 G_{4,XX}) \bigl(4 G_{4} G_{4,\phi X} + G_{4,X} (G_{4,\phi} - 6 G_{4,\phi X} X)\bigr)}{3 (G_4 - 2 G_{4,X} X)^2 (G_4 -  G_{4,X} X)}\,,
\end{equation}

 \subsubsection{DBI-like quartic theories}
\label{sec:DBI_Debraided}
 We report here the detailed expressions for the DBI-like quartic theories introduced in \ref{sec:DBI_galileon}.
 \begin{eqnarray} 
 \mathcal{G}_0 &=& G_{2,X}+ \nonumber \\ 
&& \Bigl[8 G_2 \Lambda^4 \bigl(A(\phi)\bigr)^3 \mathcal{D} X^2 + M_{\text{Pl}}^2 \Lambda^8 \mathcal{D}^4 X \Bigl(-13 X \bigl(A^\prime(\phi))^2 + 4 \Lambda^4 \mathcal{D}^2 A^{\prime\prime}(\phi)\Bigr) \nonumber \\ 
&& + 4 \bigl(A(\phi)\bigr)^2 \Bigl(-5 M_{\text{Pl}}^2 X^4 \bigl(A^\prime(\phi))^2 + 2 \Lambda^4 \mathcal{D}^2 X \bigl(G_2 \Lambda^4 \mathcal{D} + 2 M_{\text{Pl}}^2 X^2 A^{\prime\prime}(\phi)\bigr)\Bigr) \nonumber \\ \nonumber
 && + 2 \Lambda^4 A(\phi) \mathcal{D}^2 \Bigl(-14 M_{\text{Pl}}^2 X^3 \bigl(A^\prime(\phi))^2 + \Lambda^4 \mathcal{D}^2 \bigl(G_2 \Lambda^4 \mathcal{D} + 8 M_{\text{Pl}}^2 X^2 A^{\prime\prime}(\phi)\bigr)\Bigr)\Bigr]\times \\
&& \times \Bigl(2 \Lambda^8 \mathcal{D}^3 \bigl(\Lambda^4 \mathcal{D}^2 + 2 A(\phi) X\bigr)^2\Bigr)^{-1}\,,
 \end{eqnarray}
 \begin{eqnarray}
 \mathcal{S}_0 &=& -G_{2,XX} +\nonumber \\
&& \Bigl[4 \bigl(A(\phi)\bigr)^3 \Bigl(2 G_{2,X} \Lambda^8 \mathcal{D}^3 X^2 - 3 M_{\text{Pl}}^2 X^4 \bigl(A^\prime(\phi)\bigr)^2\Bigr) + 2 M_{\text{Pl}}^2 \Lambda^{12} \mathcal{D}^6 \Bigl(-10 X \bigl(A^\prime(\phi)\bigr)^2 \nonumber \\ 
&& + \Lambda^4 \mathcal{D}^2 A^{\prime\prime}(\phi)\Bigr) + 4 \Lambda^4 \bigl(A(\phi)\bigr)^2 \mathcal{D}^2 X \Bigl(-13 M_{\text{Pl}}^2 X^2 \bigl(A^\prime(\phi)\bigr)^2 + 2 \Lambda^4 \mathcal{D}^2 \bigl(G_{2,X} \Lambda^4 \mathcal{D} \nonumber \\ 
&& + M_{\text{Pl}}^2 X A^{\prime\prime}(\phi)\bigr)\Bigr) + \Lambda^8 A(\phi) \mathcal{D}^4 \Bigl(-59 M_{\text{Pl}}^2 X^2 \bigl(A^\prime(\phi)\bigr)^2 + 2 \Lambda^4 \mathcal{D}^2 \bigl(G_{2,X} \Lambda^4 \mathcal{D} \nonumber \\ 
&& + 4 M_{\text{Pl}}^2 X A{\prime\prime}(\phi)\bigr)\Bigr)\Bigr]\Bigl(2 \Lambda^{12} \mathcal{D}^5 \bigl(\Lambda^4 \mathcal{D}^2 + 2 A(\phi) X\bigr)^2\Bigr)^{-1}\,,
 \end{eqnarray}
 \begin{eqnarray}
 \tilde V &=& G_{2,\phi}-2 G_{2,\phi X}X+ \nonumber \\
 && \Bigl[X A^{\prime}(\phi) \biggl(\Lambda^4 A(\phi) \mathcal{D}^2 X \Bigl(-24 M_{\text{Pl}}^2 X^3 \bigl(A^{\prime}(\phi)\bigr)^2 + \Lambda^4 \mathcal{D}^2 \bigl(\Lambda^4 \mathcal{D} (7 G_2 + 10 G_{2,X} X)\,, \nonumber \\ 
&& - 24 M_{\text{Pl}}^2 X^2 A^{\prime\prime}(\phi)\bigr)\Bigr) + \Lambda^8 \mathcal{D}^4 \Bigl(-9 M_{\text{Pl}}^2 X^3 \bigl(A^{\prime}(\phi)\bigr)^2 + \Lambda^4 \mathcal{D}^2 \bigl(\Lambda^4 \mathcal{D} (2 G_2 + 5 G_{2,X} X) \nonumber \\ 
&& - 9 M_{\text{Pl}}^2 X^2 A^{\prime\prime}(\phi)\bigr)\Bigr) + 2 \bigl(A(\phi)\bigr)^2 X^2 \Bigl(-6 M_{\text{Pl}}^2 X^3 \bigl(A^{\prime}(\phi)\bigr)^2 + \Lambda^4 \mathcal{D}^2 \bigl(\Lambda^4 \mathcal{D} (G_2 + 4 G_{2,X} X) \nonumber \\ 
&& - 6 M_{\text{Pl}}^2 X^2 A^{\prime\prime}(\phi)\bigr)\Bigr)\Bigr]\Bigl(\Lambda^{12} \mathcal{D}^5 \bigl(\Lambda^4 \mathcal{D}^2 + 2 A(\phi) X\bigr)^2\biggr)\Bigr)^{-1}\,,
 \end{eqnarray}
 \begin{equation}
 \mathcal{V}_{4D}= \frac{2 M_{\text{Pl}}^2 A(\phi) A^\prime(\phi)}{\Lambda^8 \mathcal{D}^3}\,,
 \end{equation}
 \begin{equation}
 \mathcal{V}_{4B}=- \frac{M_{\text{Pl}}^2 \bigl(3 \Lambda^4 \mathcal{D}^2 - 2 A(\phi) X\bigr) A^\prime(\phi)}{2 \Lambda^8 \mathcal{D}^3}\,,
 \end{equation}
 \begin{equation}
 \mathcal{Q}_T = \frac{X \bigl(3 \Lambda^4 \mathcal{D}^2 + 2 A(\phi) X\bigr) A^\prime(\phi)}{2 \Lambda^8 \mathcal{D}^4 + 4 \Lambda^4 A(\phi) \mathcal{D}^2 X}\,,
 \end{equation}
 \begin{equation}
 \mathcal{Q}_{\enangle{T}} = \frac{\Bigl(4 \Lambda^8 \mathcal{D}^4 + 9 \Lambda^4 A(\phi) \mathcal{D}^2 X + 6 \bigl(A(\phi)\bigr)^2 X^2\Bigr) A^\prime(\phi)}{2 \Lambda^4 \bigl(\Lambda^4 \mathcal{D}^3 + 2 A(\phi) \mathcal{D} X\bigr)^2}\,,
 \end{equation}
 \begin{equation}
 \mathcal{C}_T = \frac{A(\phi)}{\Lambda^4 \mathcal{D}^2}\,,
 \end{equation}
 where we have defined
 \begin{equation}
 \mathcal{D} = \left(1-2A(\phi)X/\Lambda^4\right)^{1/2}\,,
 \end{equation}

\section{Existence of an Einstein frame} \label{sec:no_einstein_frame}

In section \ref{sec:weyl} we have discussed how, the presence of the Weyl tensor in the equation of motion for the scalar field can be related to the lack of a field transformation able to cast the kinetic term for the metric into its standard Einstein--Hilbert form.
In this appendix we provide additional arguments in support of this statement. More specifically we examine the requirements for a field redefinition to be able to cast a quartic theory as Einstein--Hilbert plus a coupling to matter. In what follows we enumerate different possibilities, and argue that they are implausible to produce the desired outcome.
\begin{enumerate}
 \item If we use a scalar field redefinition \emph{only} we need $G_4(X,\phi)\rightarrow M_p^2/2=\text{constant}$. However, we can see already in the old-school case that demanding that $G_4(\phi(\xi))=\text{constant}$ implies $G_{4,\phi}\frac{\partial\phi}{\partial\xi}=0$ (by taking the derivative wrt the field). This requires that either $G_4$ is constant or that the Jacobian of the field transformation is degenerate. It is hard to imagine how adding $X$ dependence would help with this.
 \item If we use a metric redefinition we note that
 \begin{enumerate}
  \item The Riemann tensor transforms as
  \begin{equation}\label{riemmangen}
  \bar R^\alpha_{\phantom{\alpha}\beta\mu\nu} =
  R^\alpha_{\phantom{\alpha}\beta\mu\nu}
  + 2\nabla_{[\mu}\mc K^{\alpha}_{\phantom{\alpha} \nu]\beta}
  + 2\mc K ^\alpha_{\phantom{\alpha} \gamma [\mu} \mc K^\gamma _{\phantom{\alpha} \nu]\beta}\,, 
  \end{equation}
  where $\mathcal K\ud{\alpha}{\mu\nu} = \bar \Gamma\ud{\alpha}{\mu\nu} -\Gamma\ud{\alpha}{\mu\nu}$ \cite{Zumalacarregui:2013pma}. So any factor that cancels the coefficient of $R$ has to come from either $\sqrt{-\bar g}$ or from $\bar g^{\beta\nu}$.
  \item Using $X$ dependent disformal transformations can neutralize all the dependences of $G_4(X,\phi)$, but non-Horndeski terms are generated in the transformation \cite{Bettoni:2013diz}. These do not lead to an unhealthy theory (see appendix \ref{sec:BH_examples}), but does not provide a canonical kinetic term for gravity either.
  Restricting to $\phi$-dependent disformal transformations prevents the non-Horndeski terms from appearing, but then only DBI-like forms of $G_4$ can be canceled. 
  
  \item If we try to use higher derivatives for the metric redefinition $\bar g_{\mu\nu}(\partial\partial\phi,\cdots...)$ then we will introduce these higher derivatives in $G_4$ \emph{unless every dangerous term cancels}. This can be seen from the structure of (\ref{riemmangen}): The derivative dependence from $\bar g^{\alpha\beta}$ and $\sqrt{-\bar g}$ will appear in $G_{4}$, unless the connection terms produce a coefficient proportional to $R$ that cancels all the field dependedences in $G_4$, with the exception of a constant term and without introducing other curvature terms. Moreover, one would in general introduce higher derivatives via the second term in \ref{riemmangen}, that also need to cancel.
 \end{enumerate}
\end{enumerate}
The above reasoning just argues the implausibility of finding a local field redefinition that transforms the kinetic term for quartic theories in a canonical form. It does not constitute a mathematical proof, but complements the discussion presented in \ref{sec:weyl}. In particular, our discussion does not apply to non-local field redefinitions.
 
\section{Beyond Horndeski: Examples} \label{sec:BH_examples}

In this appendix we briefly present the simplest theories beyond Horndeski, the pure conformal and pure disformal theories with derivative dependences. These theories can be formulated as Einstein--Hilbert plus a coupling to matter via a field redefinition. In this sense they are similar to the DBI-Galileons, and they provide toy examples of kinetic mixing in theories beyond Horndeski.
 
\subsection{Pure Conformal Theory}\label{sec:BH_pure_conformal}

In the relatively simple case of an $X$-dependent conformally coupled theory in the Jordan frame \cite{Zumalacarregui:2013pma} 
\begin{equation}\label{eq:LagConformal}
\Lag_{C} = \frac{\sqrt{-g}}{16\pi G} \lp \Omega^2 R + 6 \Omega_{,\al}\Omega^{,\al} \rp  + \sqrt{-g}\lp \Lag_\phi + \Lag_m \rp\,,
\end{equation}
one can write the equations of motion as
\begin{equation}\label{eq:conf_phi_eq}
 \nabla_\mu\lp \phi^{,\mu} \mc  T_{\rm K} \rp 
+ \frac{\Omega_{,\phi}}{\Omega_{,X}} \mc  T_{\rm K} 
- \half \frac{\delta \Lag_\phi }{\delta\phi }= 0\,.
\end{equation}
where the \emph{kinetic mixing factor} for a conformal relation reads
\begin{equation}\label{eq:confAcal}
 \mc  T_{\rm K} \equiv\frac{8\pi G  \Omega_{,X} T }{\Omega-2\Omega_{,X}X}\,.
\end{equation}
with $T=g^{\mu\nu}(T_{\mu\nu}^\phi + T_{\mu\nu}^m)$.
The derivative mixing enter (in the Jordan frame) as a way to remove the higher order derivatives of the scalar field in this type of theories. 

\subsection{Pure Disformal Theory}\label{sec:BH_pure_disformal}

Let us now consider the transformation of the Einstein-Hilbert Lagrangian under a disformal transformation depending on field derivatives
\begin{equation}
 \bar g_{\mu\nu}=g_{\mu\nu} + B(X)\phi_{,\mu}\phi_{,\nu}\,,
\end{equation}
whose associated barred connection is characterized by the following $\mc K\ud{\alpha}{\mu\nu}\equiv \bar\Gamma^\alpha_{\mu\nu}-\Gamma^\alpha_{\mu\nu}$ tensor \cite{Zumalacarregui:2013pma}
\begin{equation} \label{disfX}
\mc K\ud{\al}{\mu\nu} =  \tilde \ga^2 \phi^{,\al}\phi_{;\mu\nu}
-(\log B)_{,X}\tilde \ga^2\phi^{,\al}\phi^{,\sigma}\phi_{;\sigma(\mu}\phi_{,\nu)}
+ \half{B_{,X}}\phi_{,\mu}\phi_{,\nu}\lb \phi_{,\sigma}\phi^{;\sigma\al} - \tilde \ga^2\phi^{,\al}\enangle{\Phi}\rb\,,
\end{equation}
We will denote $\tilde \gamma^2 = B\gamma^2 = B/(1-2BX)$, $\langle \Phi\rangle = \phi^{,\mu}\phi_{;\mu\nu}\phi^{,\nu}$, $\langle \Phi^2\rangle = \phi^{,\mu}\phi_{;\mu\nu}\phi^{;\nu\sigma}\phi_{,\sigma}$, $[\Phi]=\Box\phi$, $[\Phi^2]=\phi_{;\mu\nu}\phi^{;\mu\nu}$ and $\langle R\rangle = \phi^{,\mu}R_{;\mu\nu}\phi^{,\nu}$.
The first term was already present in the only field-dependent disformal transformation, while the last two arise from the derivatives of $B$. The bulk contribution to the Einstein-Hilbert action can be computed using Eq. (38) of \cite{Zumalacarregui:2013pma}, yielding
\begin{equation}\label{EHforBX0}
 \sqrt{-\bar g}\bar R = \sqrt{-g}\lp \frac{1}{\gamma}R - B\gamma\enangle{R} - B(B+B_{,X}X)\ga^3\lp \enangle{\Phi}[\Phi]-\enangle{\Phi^2}\rp\rp\,.
\end{equation}
This simple result is due to the particular tensor structure of the $\mc K$ tensor, which makes that only the first term in Eq. (\ref{disfX}) contributes to $\bar g^{\mu\nu} \mc K ^\al_{\ph \gamma [\al} \mc K^\gamma _{\ph \mu]\nu}$ (all other contractions are proportional to a contraction of $\phi^{,\al}\phi^{,\mu}$ with a tensor antisymmetric on the indices $\al\mu$). 
The addition of a surface term $\nabla_\mu\big( f(X)( \phi^{,\mu}[\Phi] - \phi^{;\mu\al}\phi_{,\al})\big)$ to (\ref{EHforBX0}) gives
\begin{eqnarray}\label{EHforBX1}
 \sqrt{-\bar g}\bar R &=& \sqrt{-g}\Big( \frac{1}{\gamma}R -(\gamma B + f)\enangle{R} + f ([\Phi]^2-[\Phi^2]) 
 \nonumber \\ &&
 - \lp B(B+B_{,X}X)\ga^3  + f_{,X}\rp \lp \enangle{\Phi}[\Phi]-\enangle{\Phi^2}\rp\Big)\,.
\end{eqnarray}
It is not possible to write down the above action in the canonical Horndeski form, as that would require to simultaneously satisfy
\begin{enumerate}
 \item $f= (1/\ga)_{,X}=-\ga B - \ga B_{,X} X$, to have the right coefficient of $[\Phi]^2-[\Phi^2]$,
\item $f_{,X} = -B(B+B_{,X}X)\ga^3$, to kill the last term and
\item  $f=-\ga B$, to kill the $\enangle{R}$ term.
\end{enumerate}
The last choice gives the simplest form for the action
\begin{equation}\label{eq:disf_beyond_LH}
\sqrt{-\bar g}\bar R = \sqrt{-g}\left(\frac{1}{\gamma}R - B\gamma ([\Phi]^2-[\Phi^2]) 
 - \gamma B_{,X} \lp \enangle{\Phi}[\Phi]-\enangle{\Phi^2}\rp\right)\,
\end{equation}

Similarly to the conformal theory, the equations of motion for the theory (\ref{eq:disf_beyond_LH}) contain an implicit constraint that allows one to remove the higher derivatives in the equations of motion. In this case the kinetic mixing factor reads
\begin{equation}
 \mathcal T_K^D = 8\pi G \gamma \frac{B_{,X}\phi_{,\mu}T^{\mu\nu}_{\rm tot}\phi_{,\nu}}{1+2B_{,X}X^2}\,,
\end{equation}
and also enters the field equation as $\bar \nabla_\alpha (\mathcal T_K^D \phi^{,\alpha})$ \cite{Zumalacarregui:2013pma}.

\bibliographystyle{h-physrev}
\bibliography{disformal}
 
\end{document}